\documentclass[prb,aps,amsmath,amssymb,twocolumn]{revtex4-2}
\usepackage{graphicx,amsmath,amssymb,color,xcolor}
\usepackage[caption=false]{subfig}
\usepackage{setspace}
\usepackage[utf8]{inputenc}
\usepackage{ulem}
\usepackage{nicefrac}
\usepackage{multirow, makecell, ragged2e}
\usepackage[titletoc,title]{appendix}
\usepackage[colorlinks,bookmarks=true,citecolor=blue,lin kcolor=red,urlcolor=blue]{hyperref}
\usepackage{cancel}
\usepackage{bm}
\usepackage{orcidlink}

\newcommand{\be}{\begin{equation}}
\newcommand{\ee}{\end{equation}}

\newcommand{\ba}{\begin{eqnarray}}
\newcommand{\ea}{\end{eqnarray}}

\newcommand{\LLL}{\text{P}_{\text{LLL}}}

\newcommand{\LL}{$\Lambda$L}
\renewcommand{\vec}[1]{\mbox{\boldmath$#1$}}

\def\beq{\begin{eqnarray}}
\def\eeq{\end{eqnarray}}

\newcommand{\ket}[1]{\vert #1 \rangle}
\newcommand{\bra}[1]{\langle #1 \vert}

\begin{document}
\title{Probing fractional quantum Hall effect by photoluminescence}
\author{Aamir A. Makki, Mytraya Gattu, and J. K. Jain~\orcidlink{000-0003-0082-5881}}
\affiliation{Department of Physics, 104 Davey Lab, Pennsylvania State University, University Park, Pennsylvania 16802,USA}
\begin{abstract}
The recent discovery of fractional quantum anomalous Hall (FQAH) states—fractional quantum Hall states realized without an external magnetic field—in twisted transition-metal dichalcogenide (TMD) bilayers represents a significant development in condensed-matter physics. These states were first observed via photoluminescence (PL) spectroscopy. Surprisingly, a general theoretical understanding of PL is not available even for the conventional fractional quantum Hall (FQH) states. Our study of a fully spin-polarized system brings out the following picture. At the Jain filling factors $\nu=n/(2n\pm 1)$, when an electron is excited from the valence to the conduction band, the photo-excited electron and the hole lower their energies by turning into composite-fermion (CF) particles and holes. These are charge $\mp e/(2n\pm 1)$ anyons and, in general, form a bound CF exciton with energy below the band gap. Away from $\nu=n/(2n\pm 1)$, the CF exciton either binds with or annihilates an already present anyon (i.e. a CF in a partially filled level) to form a trion or a single CF quasiparticle, further lowering the energy. When the interaction between two electrons is independent of whether they occupy the conduction or the valence band, as in an ideal two-dimensional system, the resulting SU(2) symmetry implies that the energy of the emitted photon must equal the band gap, independent of correlations. In this case, the bound CF excitons and CF trions are dark, indicating an absence of photoemission at zero temperature, except for the exciton of the $\nu=1/3$ ($\nu=2/3$) state in the valence (conduction) band, but we show that the temperature dependence of the PL peak intensity can be used to measure their binding energies. We also study, using exact diagonalization methods, the effect of an SU(2)-symmetry breaking term; we find that as the strength of this term is increased, the dominant peaks of the SU(2) symmetric theory evolve continuously, while additional, low-intensity peaks appear. We discuss implications for PL experiments in semiconductor quantum wells and twisted TMD bilayers.
\end{abstract}

\maketitle

\section{Introduction}\label{sec:intro}

\begin{figure}[t]
\includegraphics[width=0.49\textwidth]{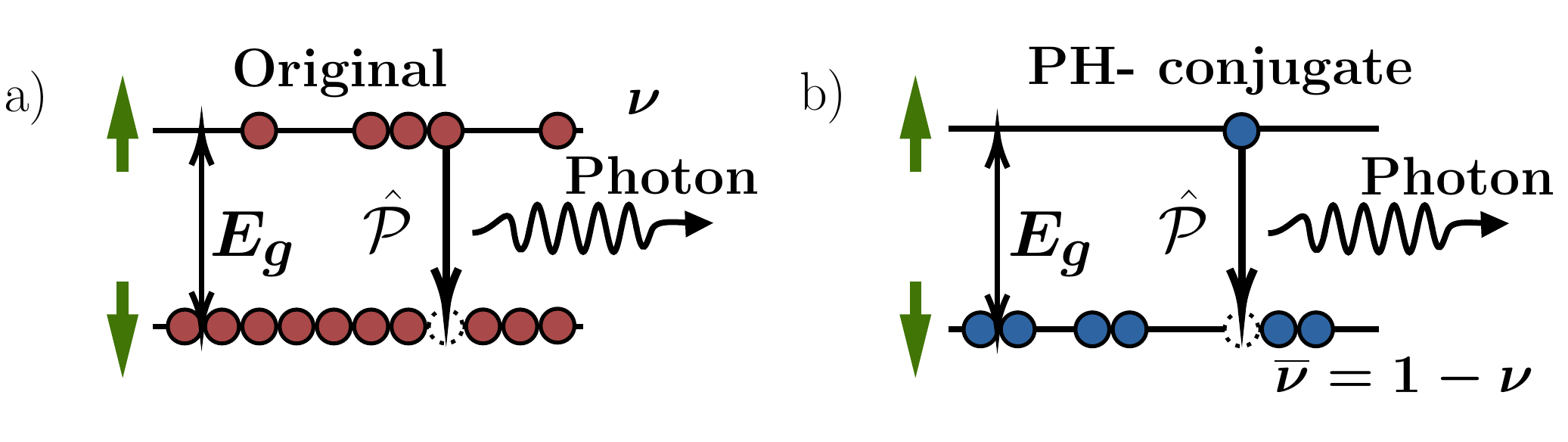}
\caption{(a) The state with a hole in LL$_\downarrow$. The electron-hole recombination is associated with the emission of a photon, which produces the PL spectrum. (b) The equivalent system related by particle-hole transformation (electron $\Leftrightarrow$ hole; $\uparrow \Leftrightarrow \downarrow$, $E\rightarrow -E$). (Note that replacing $\nu$ by $1-\nu$ in LL$_{\uparrow}$ of the left panel does not produce an equivalent system.)
\label{fig:Schematic2}
}
\end{figure}

Photoluminescence (PL) typically occurs due to the recombination of electrons in the conduction band with photoexcited holes in the valence band. 
Many experimental papers have reported PL studies of two dimensional electrons in the presence of a strong magnetic field, when the electrons in the conduction band are in the fractional quantum Hall (FQH) regime~\cite{Turberfield90,Goldberg88,Goldberg90,Byszewski06,Blokland10,Bellani10,Berezhnoy24}.
Theoretical investigation of PL in the FQH regime is primarily limited to exact diagonalization (ED) on systems containing a few (typically less than 10) electrons~\cite{Apalkov92,Rashba93,Apalkov95,Portnoi96,Macdonald92,Chen94,Wojs00a,Wojs00b,Wojs06a}, which are too small to represent the thermodynamic behavior. Furthermore, only $\nu=1/3$ has been treated in quantitative detail~\cite{Wojs06a,Wojs00b,Wojs00a}. Given that PL \cite{Cai23} and reflectance \cite{Zeng23} spectroscopy has been used to identify fractional quantum anomalous Hall (FQAH) states at several Jain fractions as well as the CF Fermi liquid \cite{Anderson24} in twisted transition-metal dichalcogenide (TMD) bilayers in a zero magnetic field, it is all the more important to gain a better understanding of the physics of the PL in the context of the standard FQH effect (FQHE) in a magnetic field. Our objective in this article is to address the issue using the composite fermion (CF) theory~\cite{Jain89,Jain07,Halperin20}, which allows us to obtain the thermodynamic behavior over a wide range of filling factors while, at the same time, revealing the underlying physics in an intuitive fashion. We show that the PL experiments can not only help identify FQHE states but also serve as a probe of CF excitons and CF trions that form at general fillings.

\section{Spin Model for Photoluminescence}\label{sec:spin_model}

We refer to the Landau levels (LLs) in the conduction and valence bands as the pseudospin-$\uparrow$ and pseudospin-$\downarrow$ LLs (LL$_\uparrow$ and LL$_{\downarrow}$), and the bandgap $E_g$ separating them as the pseudo-Zeeman splitting (Fig.~\ref{fig:Schematic2}). We assume that the real spin is locked in the favorable direction, and thus does not play any role.  The PL intensity $P(\omega)$ is given by \cite{Macdonald92,Chen94}
\begin{equation}
P(\omega)\propto\sum_{\alpha,i}e^{-{E^{\rm in}_i\over T}}\left|\bra{\phi^{\rm f}_\alpha}\mathcal{\hat P}\ket{\psi^{\rm in}_i}\right|^2\delta(\omega - (E^{\rm in}_i+E_g - E^{\rm f}_\alpha))\label{eq:PLintensity}
\end{equation}
Here $\psi^{\rm in}_i$ are the eigenstates of the initial system containing a hole in LL$\downarrow$, and $\phi^{\rm f}_\alpha$ are the eigenstates of the system where the hole has recombined with an electron from the conduction LL, with $E^{\rm in}_i+E_g$ ($E_i^{\rm in}$ is the interaction energy) and $E^{\rm f}_\alpha$ labeling the corresponding eigenenergies. It has been assumed that the initial system relaxes through non-radiative transitions to either the ground state ($T=0$) or a thermal distribution ($T>0$) before recombination occurs. The PL operator $\mathcal{\hat P}$ is defined as
\begin{equation}
\mathcal{\hat P} = \int d^2r \psi_{\downarrow}^\dagger(\vec{r})\psi_{\uparrow}(\vec{r}) 
= \sum_m \psi_{\downarrow}^\dagger(m) \psi_{\uparrow}(m)
\label{eq:PLoperator}\end{equation}
where $\psi_{\uparrow}(\vec{r})$ destroys an electron at $\vec{r}$ while $\psi_{\uparrow}(m)$ destroys an electron in orbital $m$. Note that $\mathcal{\hat P}$, which flips the pseudospin but does not change the orbital index, is nothing but the pseudospin lowering operator. 

We  assume in most of our discussion below the ideal limit of $B\rightarrow\infty$, zero thickness and no disorder. The interaction is independent of the pseudospin, i.e. the system exhibits an $SU(2)$ symmetry within the pseudospin space. In this case, the operator $\mathcal{\hat P}$ commutes with the Hamiltonian and thus implies $E_i^{\rm in}=E^{\rm f}_\alpha$ for all nonzero matrix elements $\bra{\phi^{\rm f}_\alpha}\mathcal{\hat P}\ket{\psi^{\rm in}_i}$ in Eq.~\ref{eq:PLintensity}. Consequently, the energy of the  emitted photon, $E_g$, does not contain any information regarding the correlations in the FQHE state~\cite{MacDonald90,Macdonald92}. However, we show below that the {\it intensity} of PL can not only help identify incompressible states but also be used to probe CF exciton and CF trion bound states.

\begin{figure*}[t]
\includegraphics[width=0.8\textwidth]{fig_basis_latest.pdf}
\caption{This figure depicts the physics of photoluminescence in the FQHE in terms of CFs. We consider FQHE states in the valence band at filling factors $\bar{\nu}=n/(2n+1)$, $\bar{\nu}\gtrsim n/(2n+1)$ [shown for simplicity as the $\bar{\nu}=n/(2n+1)$ state plus a quasiparticle (QP), which is a single CF in an otherwise unoccupied $\Lambda$L] and $\bar{\nu}\lesssim n/(2n+1)$ [shown for simplicity as the $\bar{\nu}=n/(2n+1)$ state plus a quasihole (QH), which is a single missing CF in the topmost occupied $\Lambda$L]. When an electron is excited from the valence band (denoted by pseudospin-$\uparrow$) to the conduction band  (denoted by pseudospin-$\downarrow$), a very complicated and highly excited state of CFs is obtained which can be written as linear combination of CF basis functions of the type shown in Column-I  (``Pre-relaxation") which have $S_z=N/2-1$. Each basis function is shown in Dirac's $|\cdots \rangle$ (``bra-ket") notation, with the pseudospin-$\uparrow$ state shown schematically on the left and the pseudospin-$\downarrow$ state on the right. We make the customary assumption that, at zero temperature, prior to recombination this state relaxes to the ground state within the $S_z=N/2-1$ sector, resulting in a state of the kind shown in the Column-II (``Post-relaxation"). Depending on the filling factor, this state may contain a CF QP in $\downarrow$ sector, a CF exciton or a CF trion. The states in Column-II can have $(S, S_z)=(N/2, N/2-1)$ or $(S, S_z)=(N/2-1, N/2-1)$. Given that the state after recombination necessarily has $(S, S_z)=(N/2, N/2)$ and that the photoluminescence operator $S^{+}$ increases $S_z$ by one unit but leaves $S$ invariant, one can show, using the CF theory, that the only ``bright" state is the $\bar{\nu}=1/3$ ground state; the ground states at all other $\bar{\nu}$ considered here are ``dark," i.e., do not allow recombination for symmetry reasons. The lowest energy bright state (which produces the state shown in Column-III after recombination) is an excited state. At finite temperatures the lowest energy bright state has a nonzero probability of occupation proportional to $e^{-\Delta/T}$, resulting in a PL signal; here $\Delta$ is the energy difference between the lowest energy bright state relative to the ground state, estimated in the main text.}
\label{fig:Schematic3}
\end{figure*}

In Sec.~\ref{sec:su2_broken}, we consider interactions that break the SU(2) symmetry, i.e., the interband interaction is different from the intraband interactions in exact diagonalization studies. We find that the dominant peaks of the SU(2) symmetric theory evolve continuously as the strength of the SU(2) symmetry breaking term is increased, while additional, broad low-intensity peaks appear. 

We find it convenient to work with the equivalent system shown in the right panel of Fig.~\ref{fig:Schematic2}, which is related to that shown in the left panel by particle-hole transformation. For simplicity, we will continue to refer to the particles as electrons (rather than holes). Now the state of interest has $\bar{\nu}=1-\nu$ and $S_z=N/2-1$.  Because the CF theory is known to provide a quantitatively accurate description of the low-energy eigenstates of both fully and partially polarized systems~\cite{Wu93,Park98,Nakajima94,Du95,Kukushkin99,Mandal01,Dujovne03b,Dujovne05,Kang97,Davenport12,Park00b,Wurstbauer11,Liu14,Balram15,Zhang16,Huang22}, a thermal distribution implies that during the relaxation, the electron excited to the LL$_{\downarrow}$ in Fig.~\ref{fig:Schematic2}(b) captures vortices to transform into a CF. 

\section{Review of CF Theory for Spinful systems}

We recall certain basic facts regarding CFs: strongly interacting electrons map into weakly interacting CFs that see a reduced magnetic field; CFs form their own LLs, called $\Lambda$Ls; and their filling factor $\nu^*$ is related to the electron filling factor $\bar{\nu}$ by the relation $\bar{\nu}=\nu^*/(2p\nu^*\pm 1)$, where $2p$ is the number of vortices bound to the CFs. We specialize to $2p=2$ in the following.

Let us begin by noting certain exact results for states in the sector $S_z=N/2-1$. The eigenstates of the system can be divided into two groups: 
\begin{itemize}
\item Dark eigenstates: The eigenstates with $(S,S_z)=(N/2-1,N/2-1)$ are annihilated by the $S_z$ raising operator. These do not allow electron-hole recombination and do not produce any PL signal. 
\item Bright eigenstates (also referred to as ``multiplicative states"~\cite{Chen94}): These have $(S,S_z)=(N/2,N/2-1)$. Upon application of the $S_z$ raising operator, each bright eigenstate produces an eigenstate in the $(S,S_z)=(N/2,N/2)$ sector with the same {\it interaction} energy. Alternatively, all bright states can be obtained by applying the $S^{-}$ operator to the eigenstates in the fully polarized $(S,S_z)=(N/2,N/2)$ sector. 
\end{itemize}

Let us first consider a mean-field model that treats CFs as noninteracting, i.e., the electron-electron interaction enters entirely through the CF kinetic energy (CFKE). The lowest energy bright states at $\bar\nu=n/(2n\pm 1)$ are obtained by application of $S_z$ lowering operator $S^{-}$ to the lowest CFKE states in the fully polarized sector, as shown in the right columns of Fig.~\ref{fig:Schematic3}. The lowest CFKE states with $S_Z=N/2-1$ are shown in the middle column of Fig.~\ref{fig:Schematic3}, and, with the exception of the $\bar\nu=1/3$ state and its QHs in Fig.~\ref{fig:Schematic3} (a,d), these have lower CFKE than the lowest bright states; this already implies that the states in the middle column are dark, but that can also be proven rigorously, with the exception of the $L=0$ state at $\bar\nu=1/3$ (see Appendix ~\ref{sec:darkness}). As a result, the model of noninteracting CFs implies that the ground states at all fillings, with the exception of the $\bar\nu=1/3$ state and its QHs, are dark and should show no PL signal at $T=0$.

We test our conclusions through explicit calculations that include residual CF interactions. These are performed in the spherical geometry~\cite{Haldane83}, where electrons move on the surface of a sphere subjected to a radial magnetic field of $2Q\phi_0$ flux quanta ($\phi_0 = h/e$). The single-particle states are monopole harmonics $Y_{Qlm}(\Omega)$, eigenstates of $L^2$ and $L_z$ with $l = Q + n_l$ and $m = -l, \dots, l$; $\Omega = (\theta, \phi)$ denotes the electron coordinates, and $n_l = 0, 1, \dots$ labels Landau levels. For spinful electrons with translationally invariant interactions, the many-body eigenstates are labeled by total angular momentum $L^2$ and total spin $S^2$, as well as their projections $L_z$ and $S_z$. We use the same notation for single-particle and total angular momentum operators; the context will make their meaning clear. Length is measured in units of magnetic length $\ell = \sqrt{\hbar / eB}$ and energy in units of $e^2/\epsilon\ell$, in which the sphere’s radius is $R = \sqrt{Q}\ell$. 

CFs experience an effective flux $2Q^{\star}\phi_{0} = 2Q-2p(N-1)$. The 
wavefunction for $N$ CFs at $\bar{\nu}^{\star} = n$ ($N$ electrons at $\bar{\nu} = n/(2n+1)$) is given by $\Psi_{n/(2n+1)}={\cal P}_{\rm LLL}\Phi_n \Phi_{1}^2$. Here, $\LLL$ is the lowest LL (LLL) projection operator, $\Phi_n$ is the wave function of $n$ filled LLs at effective flux $2Q^{\star}\phi_{0}$ and $\Phi_{1}^{2}$ is the wave function of $1$ filled LL at effective flux $2Q_{1}\phi_{0}=(N-1)\phi_{0}$. The wave functions of a quasiparticle (QP) or a quasihole (QH) are obtained by replacing $\Phi_n$ by the wave function of a QP or a QH of the IQHE state. When one particle has spin $\downarrow$, the state is given by a linear superposition of basis functions (i.e. CF states) of the form
\begin{equation}
\LLL \mathcal{A}[\Phi^{\uparrow}(\Omega_1,\cdots, \Omega_{N-1})\Phi^{\downarrow}(\Omega_N) u_1 u_2 \cdots u_{N-1}d_N]\Phi_1^2
\end{equation}
where $\Phi({\Omega_1,\cdots,\Omega_{N-1}})$ is an antisymmetric Slater determinant wave function and $\Phi(\Omega_{N})$ is the wave function of a single electron at effective flux $2Q^{\star}\phi_{0}$, and $u$ and $d$ are the spinors corresponding to up and down pseudo-spins. For $\nu=n/(2n-1)$, we replace above $\Phi_n\rightarrow [\Phi_n]^*\equiv \Phi_{-n}$ at flux $2Q^*=-[n^{-1}N-n]$.
The calculation methods have been described in the literature. The projection will be carried out by the quaternion implementation~\cite{Gattu25} of the Jain-Kamilla (JK) method~\cite{Jain97,Jain97b}. In some cases, we will employ the method of CF diagonalization (CFD)~\cite{Mandal02} to obtain more accurate energies. The accuracy of the CF results can be determined by comparing them with ED results available for small $N$; such comparisons are shown in Appendix~\ref{sec:CFD}.  

\begin{figure}[t]
\includegraphics[width=0.25\textwidth]{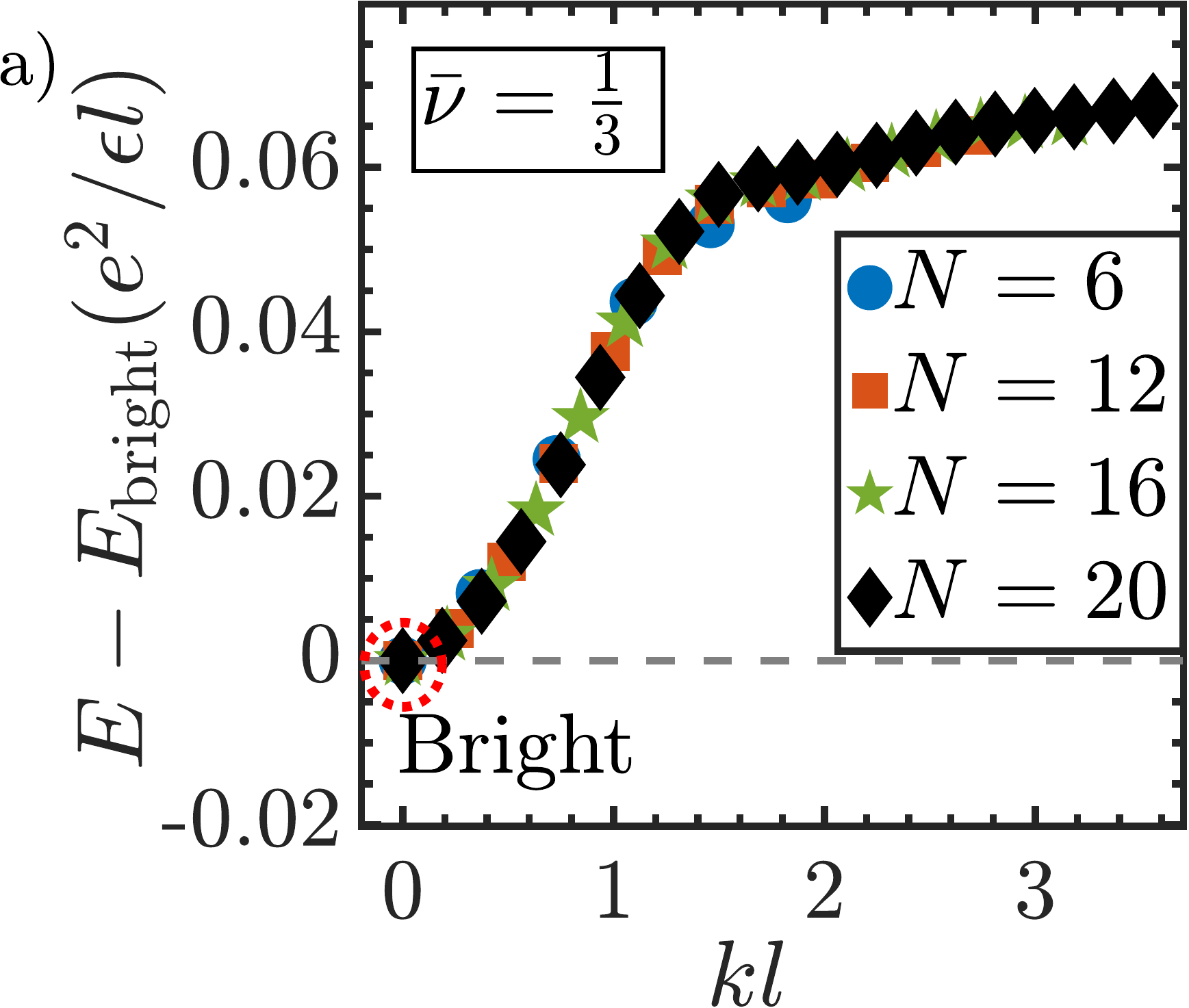}\includegraphics[width=0.245\textwidth]{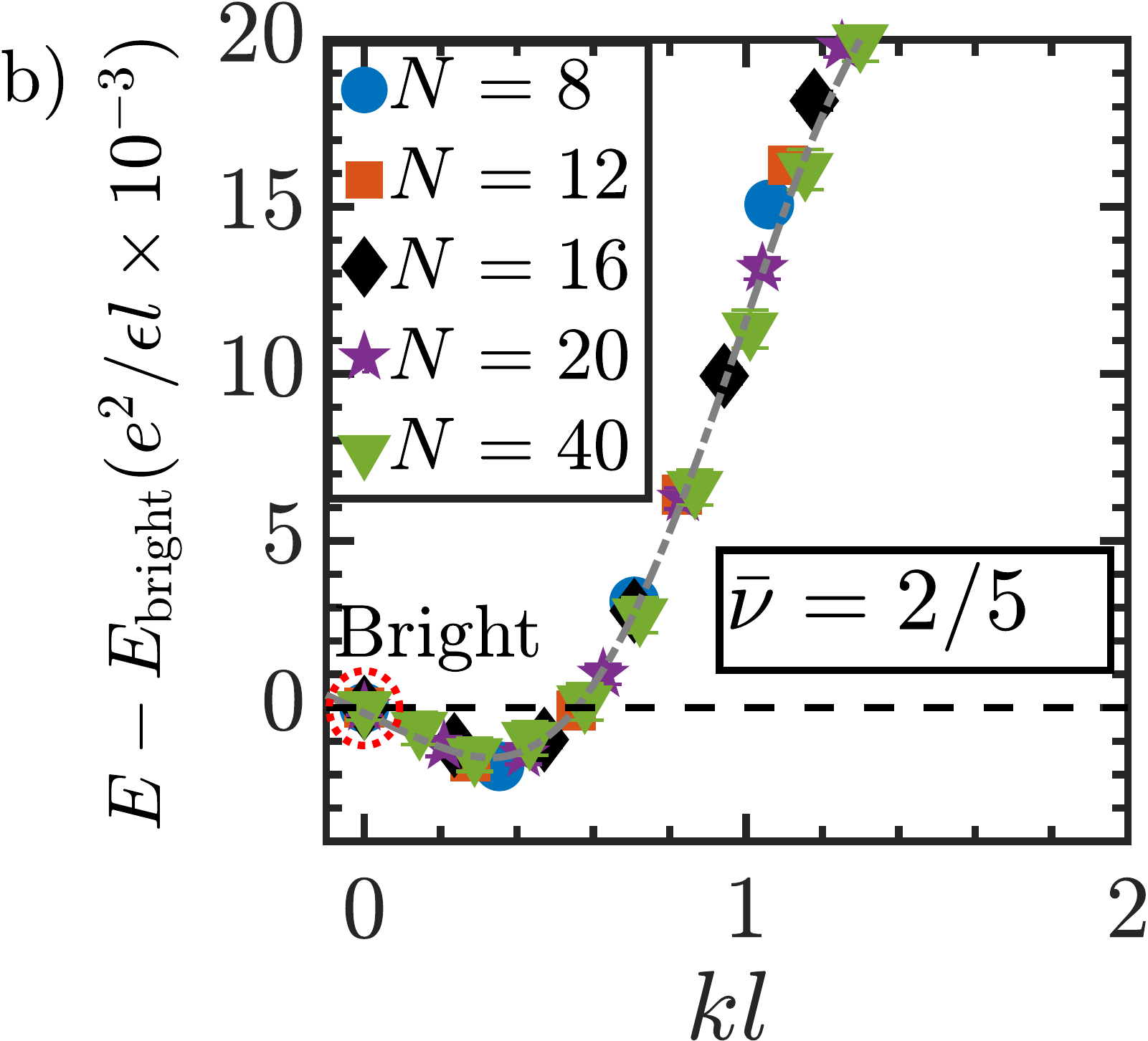}
\includegraphics[width=0.244\textwidth]{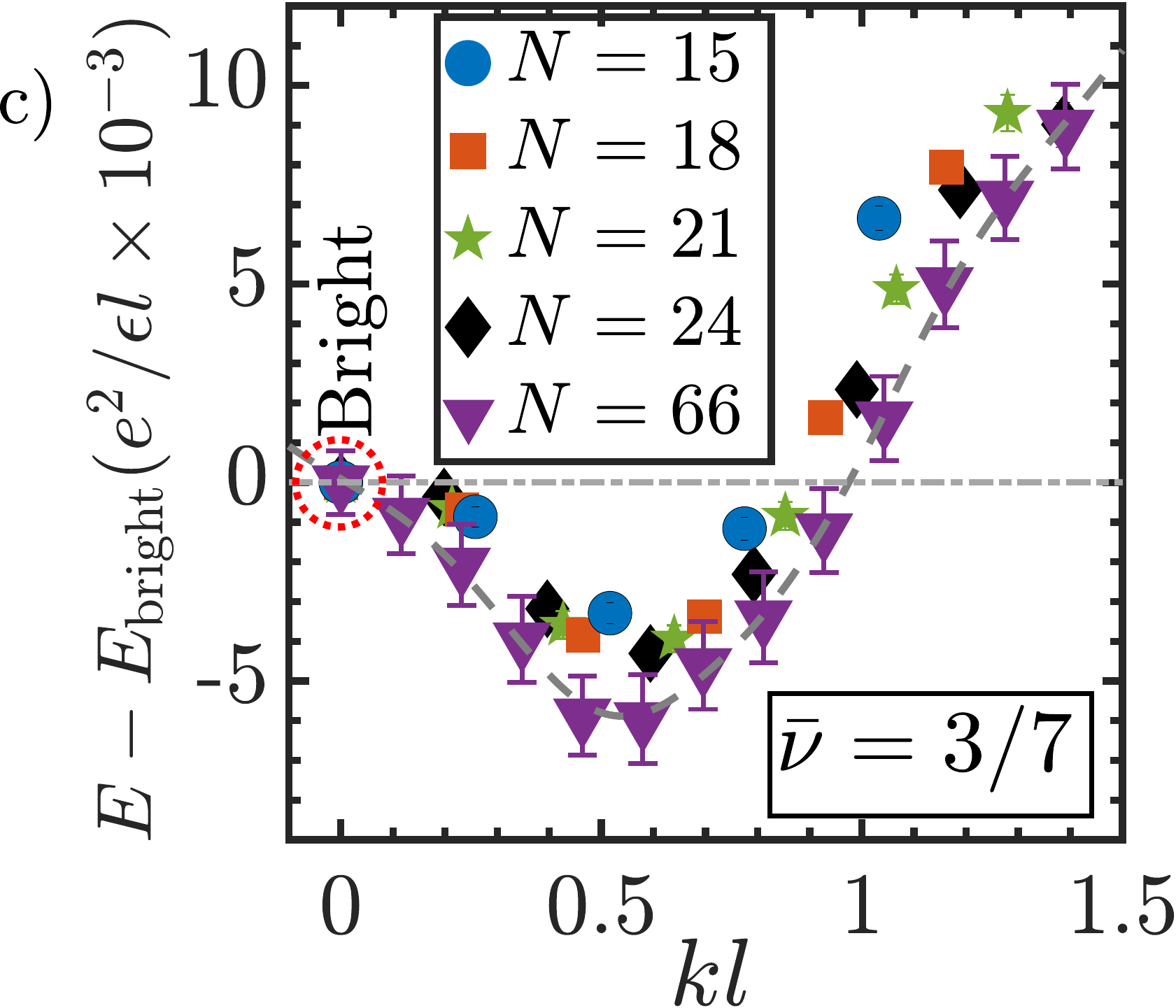}\includegraphics[width=0.25\textwidth]{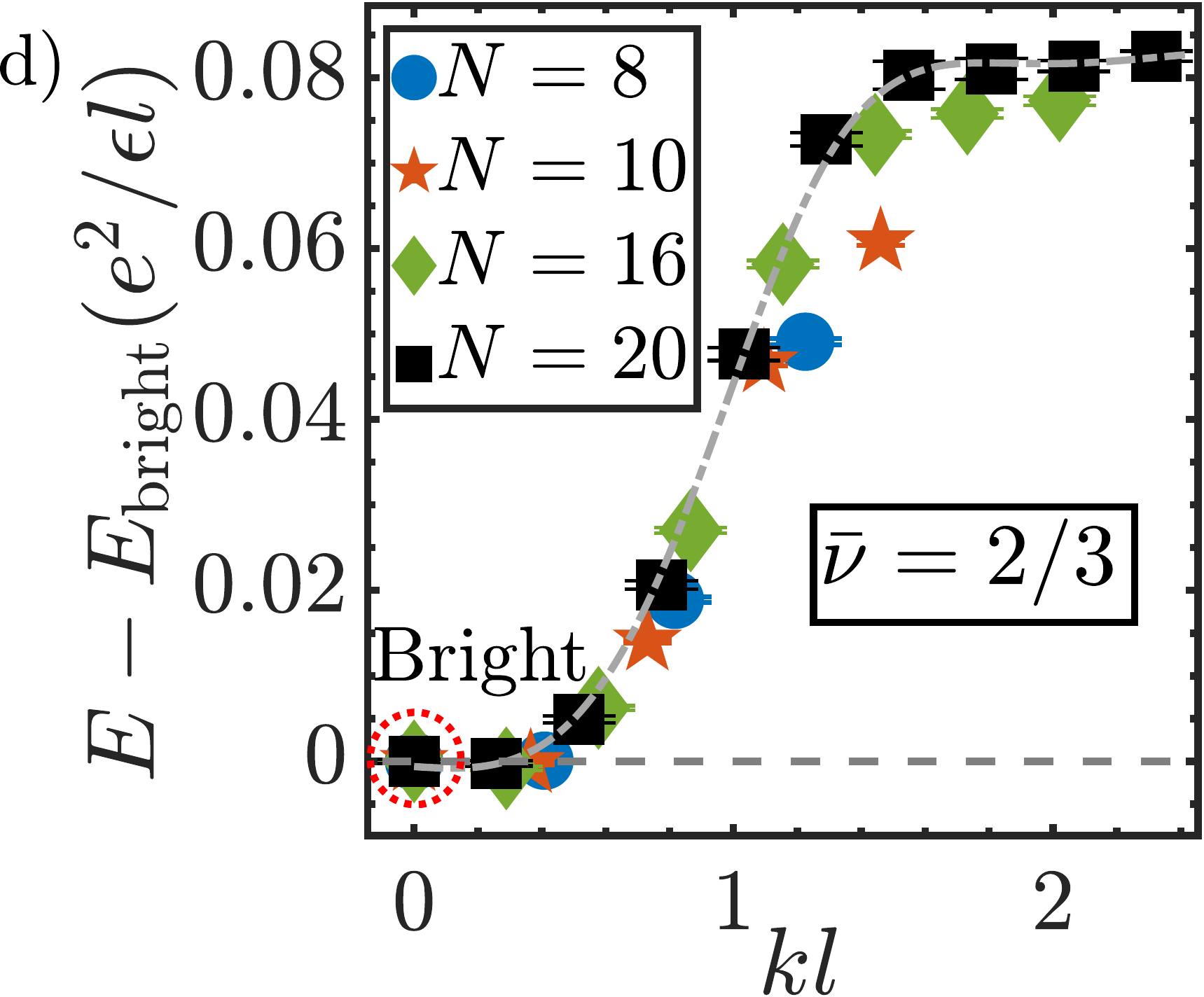}
\caption{Dispersion of the lowest pseudospin wave, obtained from CFD for: a) $\bar\nu = 1/3$, b) $\bar\nu = 2/5$, c) $\bar\nu = 3/7$, d) $\bar\nu = 2/3$. The energies are measured relative to the lowest bright state, marked by the red dashed circle; all other states shown are dark. The wave vector is defined as $k = L/R$, with $R=\sqrt{Q}l$. The pseudospin wave has a roton minimum for $\bar\nu = 2/5$ and $3/7$, producing a dark ground state.  }
\label{fig:dispersion_low_branch}
\end{figure}

\begin{figure*}[t]
\includegraphics[width=0.7\textwidth]{nu_vs_Eb_latest.pdf}
\caption{Lower panel: Plot of $\Delta$, the energy of the lowest bright state relative to the ground state, as a function of $\bar\nu$, the filling factor of the FQHE state in the valence band. The symbols are calculated [blue circle for the Jain $n/(2n+1)$ state, red square for the $n/(2n+1)$ state with a QP, green diamond for the $n/(2n+1)$ state with a QH, and the star at $\nu=1/2$]; the dashed lines are schematic. The inset shows the excitons (blue circle) and trions (green diamond and red square), where the two circles represent the $\uparrow$ and $\downarrow$ pseudospin sectors; for the red square, the QP and QH in the $\uparrow$ sector can annihilate leaving a single CF quasiparticle in the $\downarrow$ sector, as is the case for $\bar{\nu}=n/(2n+1)$ plus one QP in Fig.~\ref{fig:Schematic3}. 
Upper panel: schematic plot of the PL intensity as a function of $\bar\nu$.  
\label{fig:nu_vs_Eb}
}
\end{figure*}

\section{CF theory results}
Let us first consider $\bar\nu=n/(2n \pm 1)$. The interaction energy of the lowest bright state, $S^{-}\Psi_{n/(2n \pm 1)}$, is the same as the energy of $\Psi_{n/(2n+1)}$, and will be denoted $E^{\rm bright}_0$. One may suspect that the $S_{z}=N/2-1$ lowest energy state has a hole in the $(n-1)^{\rm th}$ $\Lambda$L$_{\uparrow}$ and a CF in the $0^{\rm th}$ $\Lambda$L$_{\downarrow}$. Whether such a state has energy $<E_{0}^{\mathrm{bright}}$ is a subtle issue because of competing contributions: the QP and the QH have positive (Hartree) energies, whereas their interaction as well as lowering of the $\Lambda$L for the QP (for $n>0$) results in a reduction of the energy. The question can only be answered by a detailed calculation. To obtain an accurate answer we consider a basis of all states where the quasihole and the quasiparticle are in the $j\uparrow$ and the  $j'\downarrow$ \LL, respectively, with $j,j'=0, 1, \cdots n-1$ being the $\Lambda$L index. Fig.~\ref{fig:dispersion_low_branch} displays the dispersion of the lowest-energy pseudospin-wave mode obtained from CFD in this basis (all modes are shown in App.~\ref{sec:CFD}). This shows that the ground state is bright for $\bar{\nu}=1/3$ and dark for $\bar{\nu}=2/5,3/7$ (because the dispersion has a roton minimum below $E_0^{\rm bright}$). The situation for $\bar\nu = 2/3$ is uncertain for the current study, with the lowest bright and dark states having the same energies within the Monte Carlo statistical error, but ED indicates that the $\bar\nu = 2/3$ state is dark for a zero width system, but finite quantum well width can brighten it (App.~\ref{sec:brighten}). Analogous studies of spin wave dispersion have been reported previously~\cite{Mandal01,Wurstbauer11}.

\begin{figure*}[t]
\includegraphics[width=0.3\textwidth]{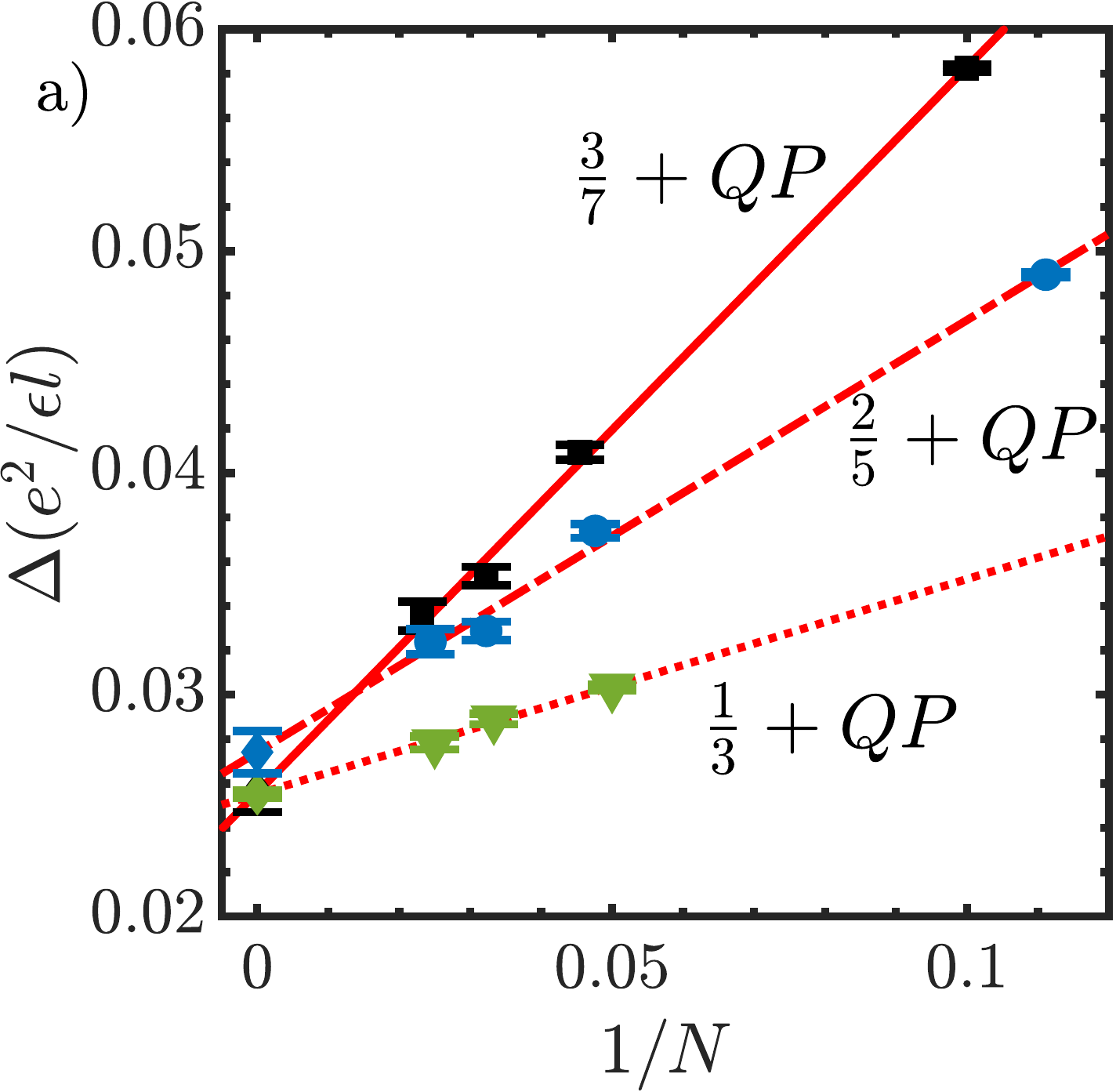}\includegraphics[width=0.3\textwidth]{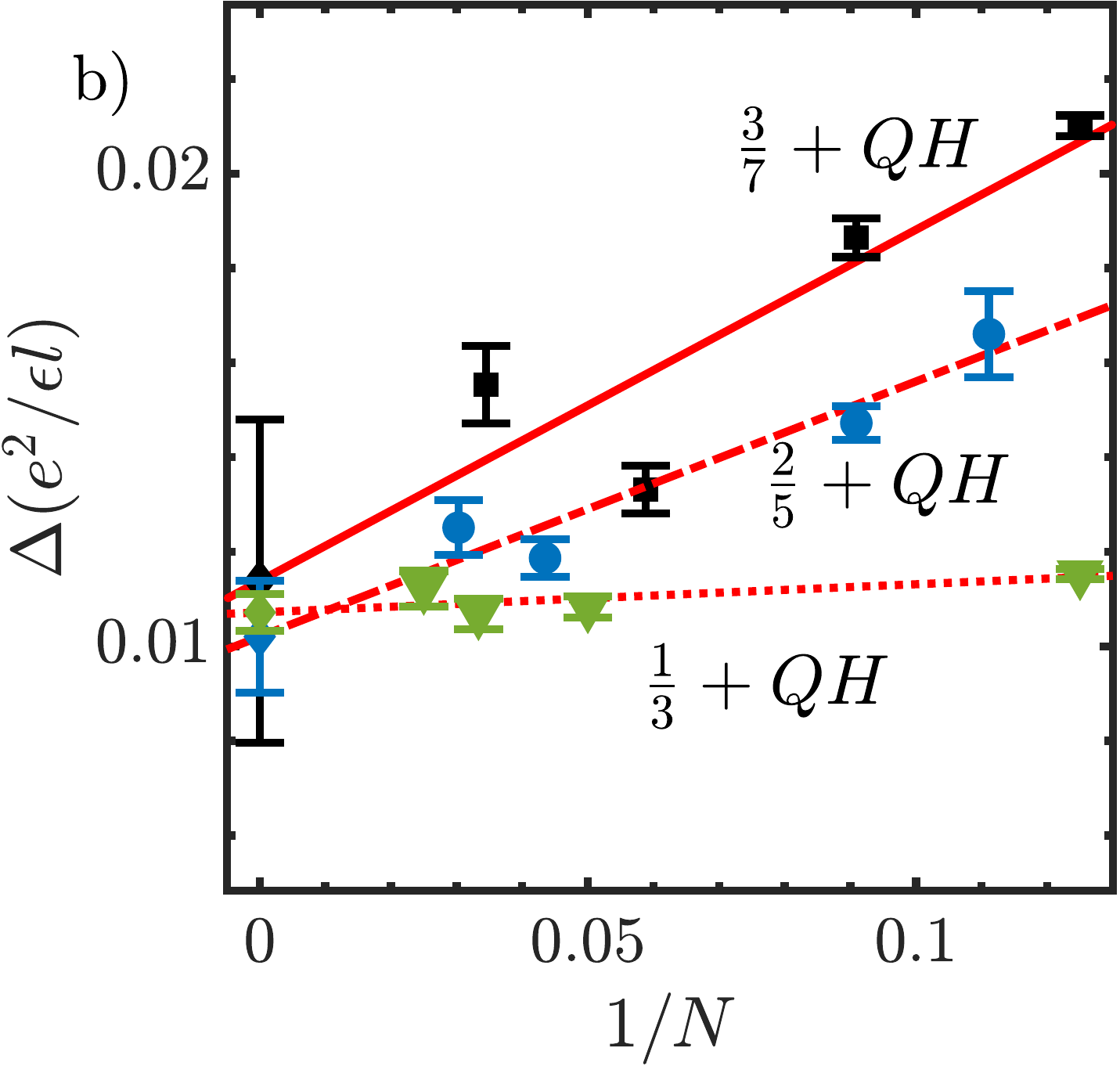}\includegraphics[width=0.3\textwidth]{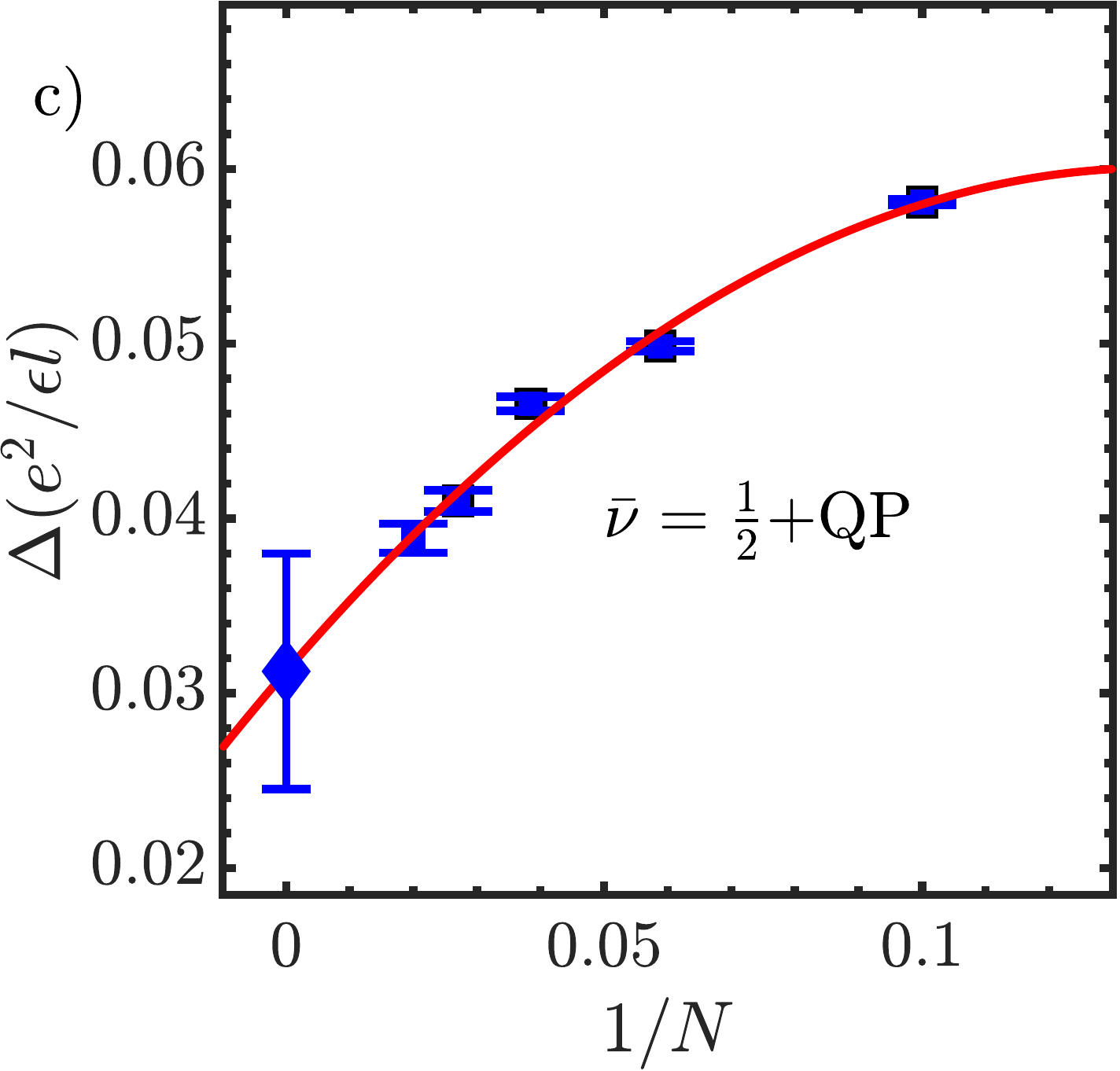}
\caption{Thermodynamic extrapolations of $\Delta$ for (a) $n/(2n+1)+{\rm QP}$, (b) $n/(2n+1)+{\rm QH}$, and (c) $1/2+\rm{QP}$. 
}\label{fig:thermo_limits}
\end{figure*}

We define $\Delta=E_0^{\rm bright}-E_0$, where $E_0$ is the energy of the ground state. For the present case, $\Delta$ is the depth of the pseudospin roton minimum. Thermodynamic extrapolation of the minimum energy in Fig.~\ref{fig:dispersion_low_branch} produces $\Delta=0.0012(3)$ for $\bar{\nu}=2/5$, $\Delta=0.0066(3)$ for $\bar{\nu}=3/7$ shown by green circles in Fig.~\ref{fig:nu_vs_Eb}. These underestimate the ED values (See App.~\ref{sec:CFD}). For $\bar{\nu}=2/3$ CFD cannot tell, within numerical uncertainty, whether the ground state is bright or dark, but ED points to the latter (App.~\ref{sec:CFD}).

Let us next consider the state which has $n$ filled $\Lambda$Ls and a single CF in the $(n+1)^{\rm}$ $\Lambda$L$_\uparrow$, which represents $\nu \gtrsim n/(2n+1)$. In this case, it is expected that moving the already present single CF to the lowest $\Lambda$L$_\downarrow$ will produce the lowest energy initial state, which is dark. We have confirmed that this agrees with the lowest dark state in ED, in both its angular momentum quantum number and its energy. Fig.~\ref{fig:thermo_limits} depicts the thermodynamic extrapolations of $\Delta$. The extrapolated values are shown by red squares in Fig.~\ref{fig:nu_vs_Eb}.

We finally consider the states at $\bar\nu\lesssim n/(2n+1)$ corresponding to Fig.~\ref{fig:Schematic3}(d-e), which are similar to Fig.~\ref{fig:Schematic3}(a,b) except that a hole is already present in the topmost $\Lambda$L. In this case, we find that the lowest energy state in the $(S,S_z)=(N/2,N/2-1)$ sector is a trion bound state. The CF exciton of Fig.~\ref{fig:Schematic3}(a,b) binds to the QH to form a CF trion to produce a lower energy state. (More complex bound states may occur when more QHs are present; we have not considered that possibility.) For $\nu\lesssim 1/3$ this is related to the skyrmion physics, and was studied in the work by Balram {\it et al.}~\cite{Balram15d}, which quantitatively explained the sub-Zeeman energy modes observed in resonant inelastic light scattering experiments in GaAs~\cite{Gallais06,Dujovne03}. To obtain the energy of the trion at other fillings, we perform CFD within the full basis at the value of $L_z$ corresponding to the configuration in which the two QHs and the QP are placed compactly near the north pole. The energies agree well with the ED energies for small systems (App.~\ref{sec:CFD}). Fig.~\ref{fig:thermo_limits} shows extrapolations to the thermodynamic limit. 

We have also included in Fig.~\ref{fig:thermo_limits} the gap $\Delta$ for $\nu=1/2$ calculated for systems with $N=n^2+1$ at $Q^*=0$. Here, we expect the same $\Delta$ for $1/2$ ground state, $1/2$ state with a QP and the $1/2$ state with a QH, because a QP-QH pair can be created without any energetic cost in the CF Fermi liquid. 

\section{Non-SU(2) symmetric interaction}\label{sec:su2_broken}

\begin{figure}[htb]
\includegraphics[width=0.4\textwidth]{ED_spectrum_80_pct_8_16_-6_0.pdf}
\includegraphics[width=0.4\textwidth]{PL_spectrum_80_pct_8_16_-6_0.pdf}
\label{fig:80_pct_2_5}
\caption{{Top: ED spectrum at $\bar\nu = 2/5$ for an $SU(2)$ broken system with $V_{\uparrow\downarrow}/V_{\uparrow\uparrow} \equiv \kappa = 0.8 $. The red diamonds, which were previously used to mark bright states, now show eigenstates in the $S_z = N/2$ sector. The blue states as before mark the eigenstates in the $S_z = N/2 -1$ sector.
Bottom: PL spectrum for the same system. All temperatures are in units of $e^2/k_B\epsilon l$. At $T=0$, the spectrum is essentially dark, as in the $SU(2)$ symmetric case. At $T=0.005$ (order of the spin roton gap), we see a single peak, adiabatically related to the single peak of the $SU(2)$ case. At higher temperatures, where the continuum of states above the low energy band is populated in the Boltzmann distribution, we see emergence of side peaks. The largest peak comes from the transition marked by a green arrow in the top figure.} 
\label{fig:80_pct_2_5}
}
\end{figure}

We now ask to what extent our results remain valid when a moderate $SU(2)$ symmetry-breaking interaction is added. We address this issue by ED calculations. In order to break the symmetry, we use a weakened inter-band ($\uparrow\downarrow$) interaction such that $V_{\uparrow\downarrow} = \kappa V_{\uparrow\uparrow}$, with $\kappa < 1$. We assume that $V_{\uparrow\uparrow}=V_{\downarrow\downarrow}$. (Non-SU(2) symmetric interaction have recently been studied also in the context of FQH effect in bilayer graphene~\cite{Huang25}.)

With the $SU(2)$ symmetry broken, the concept of ``bright" and ``dark" states no longer remains exactly valid as the eigenstates of the Hamiltonian are not necessarily $S^2$ eigenstates. In particular, the conclusion, which follows from the $SU(2)$ symmetry, that there may only be a single peak in the spectrum (at the pseudo-Zeeman energy) is also lifted. If the $SU(2)$ breaking is small, the eigenstates are adiabatically connected to the $SU(2)$ symmetric case. Particularly, the dark states $\ket{\psi^{\rm in}_i}$ evolve into states with non-zero but small matrix element $\bra{\phi^{\rm f}_\alpha}\mathcal{\hat P}\ket{\psi^{\rm in}_i}$, which is equivalent to the statement that $S^2$ operator expectation value being close but not equal to $S(S+1)$ with $S = \frac{N}{2}-1$. Similarly, the bright states evolve into states with pseudospin close to $\frac{N}{2}$. Thus, at low temperatures, when only the low energy branch of states is occupied, we expect there to be a single dominant peak, which is now shifted from the pseudo-Zeeman energy $E_g$. 

As an illustration, Fig.~\ref{fig:80_pct_2_5} shows the PL spectra and the energy eigenstates of a small system 
at $\bar\nu = 2/5$ with $\kappa = 0.8$. The $L=0$ eigenstate is adiabatically connected to the $L=0$ bright state of the $SU(2)$ symmetric system and is nearly bright i.e. has spin $S$ expectation value close to $N/2$. The ground state is nearly dark ($\langle S\rangle\sim N/2-1$), and therefore the total PL intensity at $T=0$ is very small. At a temperature $T = 0.005 e^2/k_B\epsilon l_B$, which is comparable to the gap between the nearly bright state and the ground state, the nearly bright state is occupied significantly according to the Boltzmann distribution and gives a single peak in the PL spectrum. We have verified that this peak receives most of its weight from the transition between the nearly bright state and the FQH ground state at $\bar\nu = 2/5$ (transition shown by green arrow in Fig.~\ref{fig:80_pct_2_5}). At a further higher temperature ($T = 0.015$), as the higher energy continuum of states begins to be occupied and the low energy branch has significantly thermal occupation, we see emergence of weaker side peaks, caused by the transition from these states to the exciton states of the FQH system. 

The strongest peak shown in Fig.~\ref{fig:80_pct_2_5} evolves from the sole peak allowed in the $SU(2)$ symmetric system. The conclusions in the previous section regarding the intensity as a function of $\bar\nu$ are thus valid in reference to the intensity of this peak. The spectra at other fractions have analogous properties as shown in the appendix.

\section{Discussion and Conclusion}
To summarize our principal results, we have considered an ideal 2D system and determined the $\Delta=E_0^{\rm bright}-E_0$ at the Jain fractions $\bar\nu=n/(2n\pm 1)$ and in their immediate vicinity. Depending on $\bar\nu$, the ground state can be a CF QP, a CF exciton, or a CF trion. With the exception of $\bar{\nu}=1/3$, these have $\Delta>0$ and thus are dark. Hence a recombination by PL is forbidden at $T=0$. At finite but small $T$, the PL intensity is proportional to $e^{-\Delta/T}$. The $\Delta(\bar{\nu})$ obtained from detailed calculations (above) is plotted in Fig.~\ref{fig:nu_vs_Eb} (only the symbols have been calculated; dashed lines are guides to eye), which also plots a schematic of the PL intensity as a function of $\bar\nu$. The local minima of $\Delta_{\bar\nu}$ at $\bar{\nu}=n/(2n\pm 1)$ lead to maxima in the PL intensity here, and the discontinuous change in $\Delta$ at $\bar{\nu}=n/(2n+1)$ produces an asymmetric peak. The incompressible states can thus be identified by the maxima in $P(\bar{\nu})$. Finally, the ``binding" energy of the CF exciton or the CF trion, namely the difference between the green and blue energies in Fig.~\ref{fig:nu_vs_Eb}, can in principle be determined from the activated $T$ dependence of the intensity.

{We have also shown the robustness of the predictions made for the $SU(2)$ symmetric system to moderate symmetry-breaking perturbations. Particularly, we have shown that there is a peak in the spectrum adiabatically connected to the single peak in the $SU(2)$ symmetric system whose intensity has the same behavior as a function of filling factor.}

We next compare these predictions against experimental observations. In Ref.~\cite{Turberfield90} the PL intensity in GaAs heterojunctions shows structure that appears to be correlated with the FQHE (e.g. at ~$\frac{2}{3}$, $\frac{3}{5}$,$\frac{4}{7}$,$\frac{5}{9}$) but the extrema are slightly shifted from $\nu=n/(2n\pm 1)$. In GaAs quantum wells, peak shifts/splittings are observed near several fractions (e.g. $\frac{1}{3}$, $\frac{2}{5}$,$\frac{2}{3}$,$\frac{3}{5}$,$\frac{3}{7}$) \cite{Goldberg90,Byszewski06}. Ref.~\cite{Goldberg90} reports a minimum of PL intensity at $\frac{2}{3}$. It is possible that these deviations between theory and experiments arise either due to the neglect of finite transverse width of the wave function, which affects the form of the interaction differently in the conduction and valence bands, thereby breaking SU(2) symmetry of the interaction~\cite{Byszewski06}, or due to an incomplete thermalization prior to recombination. A quantitative investigation of these issues is left for future study. 

We next come to fractional quantum anomalous Hall effect (FQAHE) in TMDs. At first sight, it would appear that our calculations do not apply to this system, and indeed the details are different, but one may ask if certain qualitative features of our study might be applicable to these systems, given that they exhibit incompressibility at the standard Jain fractions. In these systems, recombination can occur either in the same valley that contains the FQAHE state ($\sigma^{-}$) or in the other valley ($\sigma^{+}$); our theory corresponds to the $\sigma^{-}$ configuration. Interestingly, in Ref.~\cite{Anderson24}, even though FQAHE has been identified at several Jain fractions (specifically, at hole filling factors given by $\nu_h=-2/3, -3/5, -4/7$), in the $\sigma^{-}$ configuration they see a signal only at $\nu_h=-2/3$, which corresponds to our $\bar\nu=1/3$; this is consistent with our prediction in the zero $T$ limit (the experiments are performed at $T=1.6$ K, which is small compared to the $\bar\nu=2/5$ gap of $\Delta=0.007 e^2/\epsilon l\approx 6$  K assuming an effective field of 100T and $\epsilon = 10$ (using the ED value of $\Delta$ from App.~\ref{sec:CFD}). We expect negligible signal for the anomalous CF Fermi liquid; the experiments do not see any signal for $\nu_h<-1/2$ but do for $\nu_h>-1/2$. 

Our model should apply to thin quantum wells as well to PL arising from recombination of an electron-hole pair in $n=-1$ and $n=+1$ graphene LLs which show a plethora of Jain states (see, e.g. Refs.~\cite{Amet15,Dean20}). In other cases, our model is approximate but should provide a very useful starting point for analyzing experiments and for further theoretical studies.

{\it Acknowledgments -} The work was supported in part by the National Science Foundation under Grant No. DMR-2404619. We thank Xiaodong Xu, Zhen Bi and Anubhav Anilkumar for useful discussions and acknowledge Advanced CyberInfrastructure computational resources provided by The Institute for CyberScience at The Pennsylvania State University. Exact diagonalization was performed using the DiagHam libraries~\cite{DiagHam}.

{\it Data Availability -} The data that support the findings of this article are openly available \cite{github}, embargo periods may apply
\appendix

App.~\ref{sec:details} provides technical details of the CF theory calculations. In App.~\ref{sec:darkness}, we prove that the states shown in the left panel of Fig.~\ref{fig:Schematic3} are dark, with the exception of the state at $\bar\nu = 1/3$. App.~\ref{sec:CFD} introduces the basis states for CFD, compares the quantitative predictions of composite fermion (CF) theory with exact diagonalization results, and discusses the physics of the different states present in the exact and CFD spectra. In App.~\ref{sec:brighten}, we consider whether finite quantum well width can convert a dark state into a bright one. App.~\ref{app:su2break} shows additional results for $SU(2)$ broken case.
 
\section{Details of numerical calculations}
\label{sec:details}

In this section, we present details pertaining to our numerical computation of the PL intensity $P(\omega)$ as a function of $\bar{\nu}$.

Following the discussion in the main text, any CF wavefunction, at $S_{z} = (N/2-1)$, where $N$ is the number of electrons, can generally be written as a linear superposition of basis functions of the form:
\begin{equation}\label{eq:cf-1-spin-down-definition}
\begin{split}
    \psi(\{\Omega_i\}) = 
&{\cal P}_{\rm LLL}\mathcal{A}[\Phi^{\uparrow}(\Omega_1,\cdots,\Omega_{N-1})
\\
&\times\Phi^{\downarrow}(\Omega_N) u_1 u_2 \cdots u_{N-1}d_N]\Phi_1^2.
\end{split}
\end{equation}
Here, $u_i$ and $d_i$ denote the up and down spinors of the $i^{\text{th}}$ particle, and $\mathcal{A}$ is the anti-symmetrization operator. $\Phi^{\uparrow}(\Omega_1, \dots, \Omega_{N-1})$ is a Slater determinant of $N-1$ electrons, while $\Phi^{\downarrow}(\Omega_N)$ is an orbital at the effective magnetic field of the CFs. Since $\Phi^{\uparrow}$ is already anti-symmetric under permutations of its coordinates, the operator $\mathcal{A}$ in Eq.~\ref{eq:cf-1-spin-down-definition} effectively anti-symmetrizes only between ${\Omega_1, \dots, \Omega_{N-1}}$ and $\Omega_N$. For spin-independent observables, this explicit anti-symmetrization is unnecessary, as the corresponding operator $\hat{K}$ does not alter electron spin~\cite{Jain07}.

To compute the photoluminescence intensity $P(\omega)$ [see Eq.\ref{eq:PLintensity}], we identify the low-lying excited states in the $S_z = N/2 - 1$ sector. We begin by organizing composite fermion (CF) states according to their kinetic energy, and incorporate residual interactions using CF diagonalization (CFD)~\cite{Mandal02,Jain07}, where the Hamiltonian $\hat{H}$ is diagonalized within a CF basis. Since CF states with identical angular momentum quantum numbers are typically non-orthogonal, we compute both the overlap matrix $O_{ij} = \langle \Phi_i | \Phi_j \rangle$ and Hamiltonian matrix elements $H_{ij} = \langle \Phi_i | \hat{H} | \Phi_j \rangle$ via Metropolis-Hastings Monte Carlo (with ${\Phi_i}$ denoting the CF basis). The eigenstates and energies are obtained by solving the generalized eigenvalue problem $H \xi_i = \lambda_i O \xi_i$, or equivalently by diagonalizing $O^{-1} H$ after resolving linear dependencies. To simplify this, we first project the CF basis onto $L^2$ eigenstates using Clebsch–Gordan coefficients, allowing block-diagonalization of $O^{-1} H$. Details of the CF basis used in the calculation of $P(\omega)$ are given in Sec.~\ref{sec:CFD}.

\section{Identifying dark states}
\label{sec:darkness}

In this section, we demonstrate how a CF state with $S_{z} = N/2 - 1$ and $S = N/2 - 1$ can be identified using the Fock cyclic condition~\cite{Hamermesh62, Jain07}.

We have considered various states with $S_z={N\over 2}-1$. The states with $S=N/2$ are bright and those with $S={N\over 2}-1$ are dark. We now determine the $S$ for various trial states of the form shown in Fig.~2 and show that they are dark (with the exception of $L=0$ state in Fig.~2a).

The state of spinful electrons has the general form
\begin{equation}
\Psi = \mathcal{A}[\Phi(\{\Omega_1,\cdots,\Omega_N \}) u_1 u_2 \cdots d_N].
\end{equation}
Here, $\Phi(\{\Omega_1,\cdots,\Omega_N \})$ is antisymmetric with respect to exchange of any two spin-up or spin-down coordinates, and $\mathcal{A}$ represents antisymmetrization. The $S_z$ quantum number can be read off but the $S$ quantum number requires more work. However, there is a simple criterion to ascertain if the state is a highest weight state, i.e., has $S_z=S$, because in that case the state must be annihilated by the $S_z$ raising operator $S^{+}$. (A state with $S_z=-S$ is analogously annihilated by the spin lowering operator.) This is equivalent to the condition that $\Phi(\{\Omega_1,\cdots,\Omega_N \})$ be annihilated if we attempt to further antisymmetrize a spin-down coordinate with respect to the spin-up coordinates. In other words, 
\begin{equation}
  \mathcal{A}[\Phi(\{\Omega_1,\cdots,\Omega_N \})]=0\implies  S^{+}\Psi=0 \implies S=S_z.
\end{equation}
This is called the Fock cyclic condition.

We first consider a basis function of the type given in the left panels of Fig. 1 (b), (c) and (e). Let us first take the corresponding electronic state with the form
\begin{equation}
\Psi=\mathcal{A}[\Phi(\{\Omega_1,\cdots,\Omega_N \}) u_1 u_2 \cdots u_{N-1}d_N]
\end{equation}
\begin{equation}
    \Phi(\{\Omega_1,\cdots,\Omega_N \})= \Phi_{\uparrow}(\Omega_1,\cdots,\Omega_{N-1})\Phi_{\downarrow}(\Omega_N)
\end{equation}
where $\Phi_{\uparrow}(\Omega_1,\cdots,\Omega_{N-1})$ is an antisymmetric Slater determinant wave function and $\Phi_{\downarrow}(\Omega_N)$ is the wave function of a single electron in an orbital in the LLL. Now, if we attempt to further antisymmetrize $\Omega_N$ with respect to other $\Omega$'s in the function $\Phi_{\uparrow}(\Omega_1,\cdots,\Omega_{N-1})\Phi_{\downarrow}(\Omega_N)$, it vanishes, because all LLL states are occupied in the $\Phi_{\uparrow}(\Omega_1,\cdots,\Omega_{N-1})$. This implies that the wave function has $S=N/2-1$. Let us now composite-fermionize this basis function by multiplication by $\Phi_1^2$ followed by projection into the LLL, which yields Eq.~\ref{eq:cf-1-spin-down-definition}. Multiplication by $\Phi_1^2$, which is fully symmetric with respect to the exchange of any two coordinates, preserves the Fock condition and thus also $S$ and $S_z$. It is straightforward to show that LLL projection also leaves $S$ and $S_z$ invariant~\cite{Jain07}. The states shown in Fig. 1 (b), (c) and (e) are thus all dark.

Let us next consider the wave function in Fig.~2(a) with $L_z\neq 0$. For the corresponding electronic wave function at $\nu=1$, $\Phi_{\uparrow}(\Omega_1,\cdots,\Omega_{N-1})$ is an antisymmetric Slater determinant wherein a single state of the LLL is unoccupied and $\Phi_{\Omega_N}$ is the wave function of a single electron in the LLL. Again, an attempt to antisymmetrize $\Omega_N$ with respect to the other $z$'s in the function $\Phi_{\uparrow}(\Omega_1,\cdots,\Omega_{N-1})\Phi_{\downarrow}(\Omega_N)$ annihilates it, because the hole in the $\uparrow$ sector and the particle in the $\downarrow$ sector are in different angular momentum orbitals, i.e. the orbital that the particle occupies in the $\downarrow$ sector is always occupied in the $\uparrow$ sector. Composite-fermionization keeps $S$ and $S_z$ invariant, implying that all states formed from this type of basis functions with $L_z\neq 0$, and thus with $L\neq 0$, are dark. (The above proof applies to the highest weight states, but this remains true for the entire multiplet obtained by application of spin and angular momentum lowering / raising operators.)

Next let us come to the system shown in Fig.~1(d). Here, the lowest energy bright state is given by the right panel, occurring at $L=Q^*$. However, if the two CF QHs and one CF QP shown in the left panel form a trion bound state (which they do for the Coulomb interaction), it lies below $E_0^{\rm bright}$ and is thus a dark state. This is related to the skyrmion physics, studied in detail by Balram {\it et al.}~\cite{Balram15d}, which quantitatively explained the sub-Zeeman energy modes observed in resonant inelastic light scattering experiments in GaAs~\cite{Gallais06,Dujovne03}.

\section{CF diagonalization - basis states and comparison with exact results}\label{sec:CFD}

\begin{figure}[h]
\includegraphics[width=0.95\columnwidth]{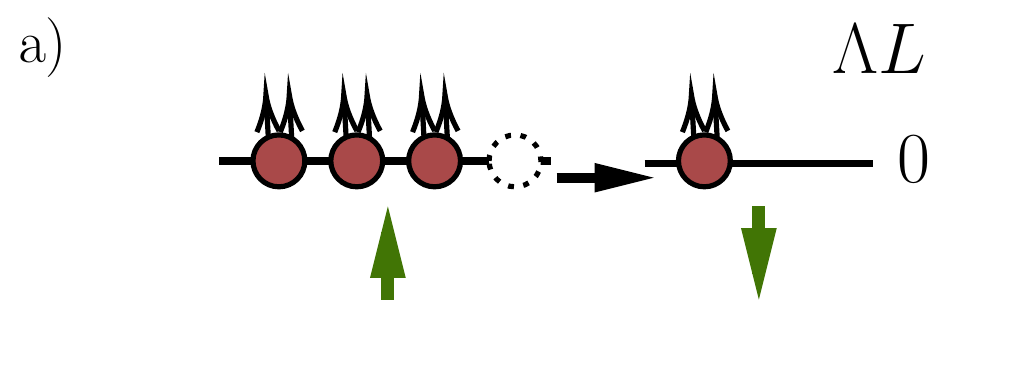}
\includegraphics[width=0.95\columnwidth]{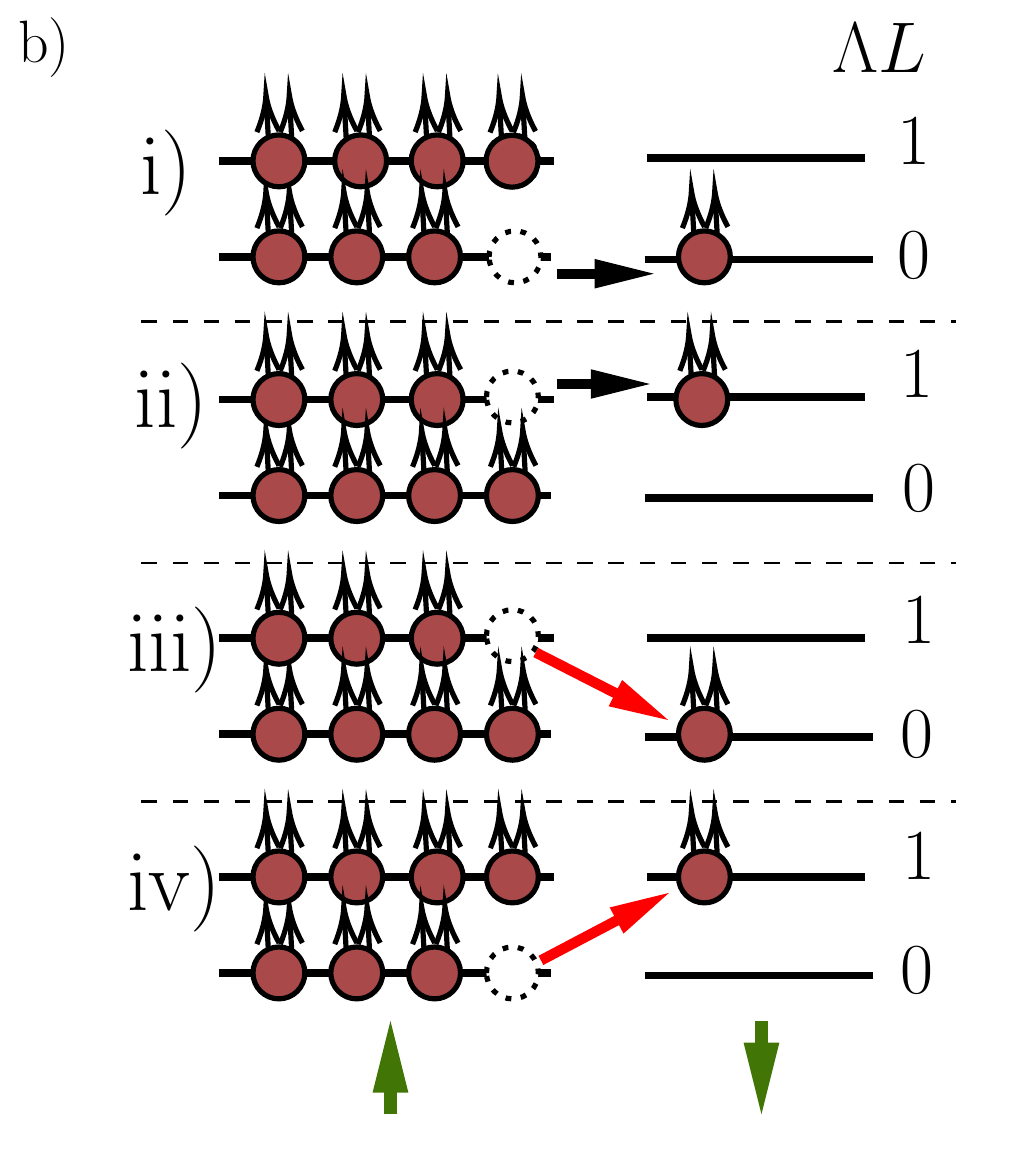}
\caption {a) The CF basis used for $\bar\nu = 1/3$. There is only a single \LL\: involved. The basis states are formed from the polarized FQHE state by moving a CF from the up band to the down band. All states have the same CFKE. b) The CF basis used for $\bar\nu = 2/5$. Those in (i) and (ii) have the same CFKE as the fully polarized ground state, while (iii) and (iv) have lower and higher CFKEs. The bright state is a linear combination of type (i) and (ii) states. Type (iii) and (iv) states are always dark. In general, for $\bar\nu = n/(2n+1)$, $n^2$ excitations are allowed. 
}\label{fig:basis_1_3_2_5}
\end{figure}

\begin{figure}[h]
\includegraphics[width=0.45\textwidth]{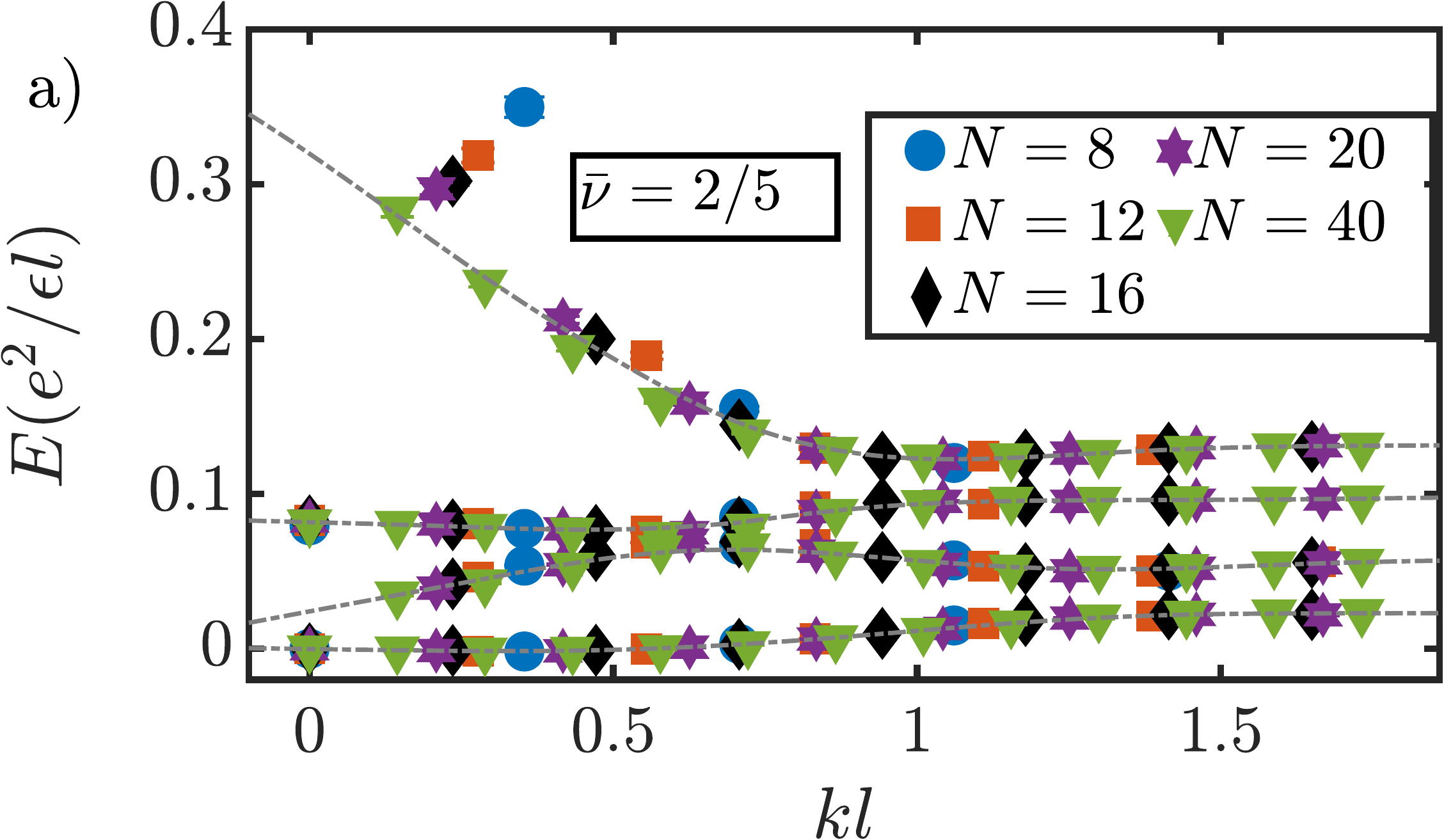}
\includegraphics[width=0.45\textwidth]{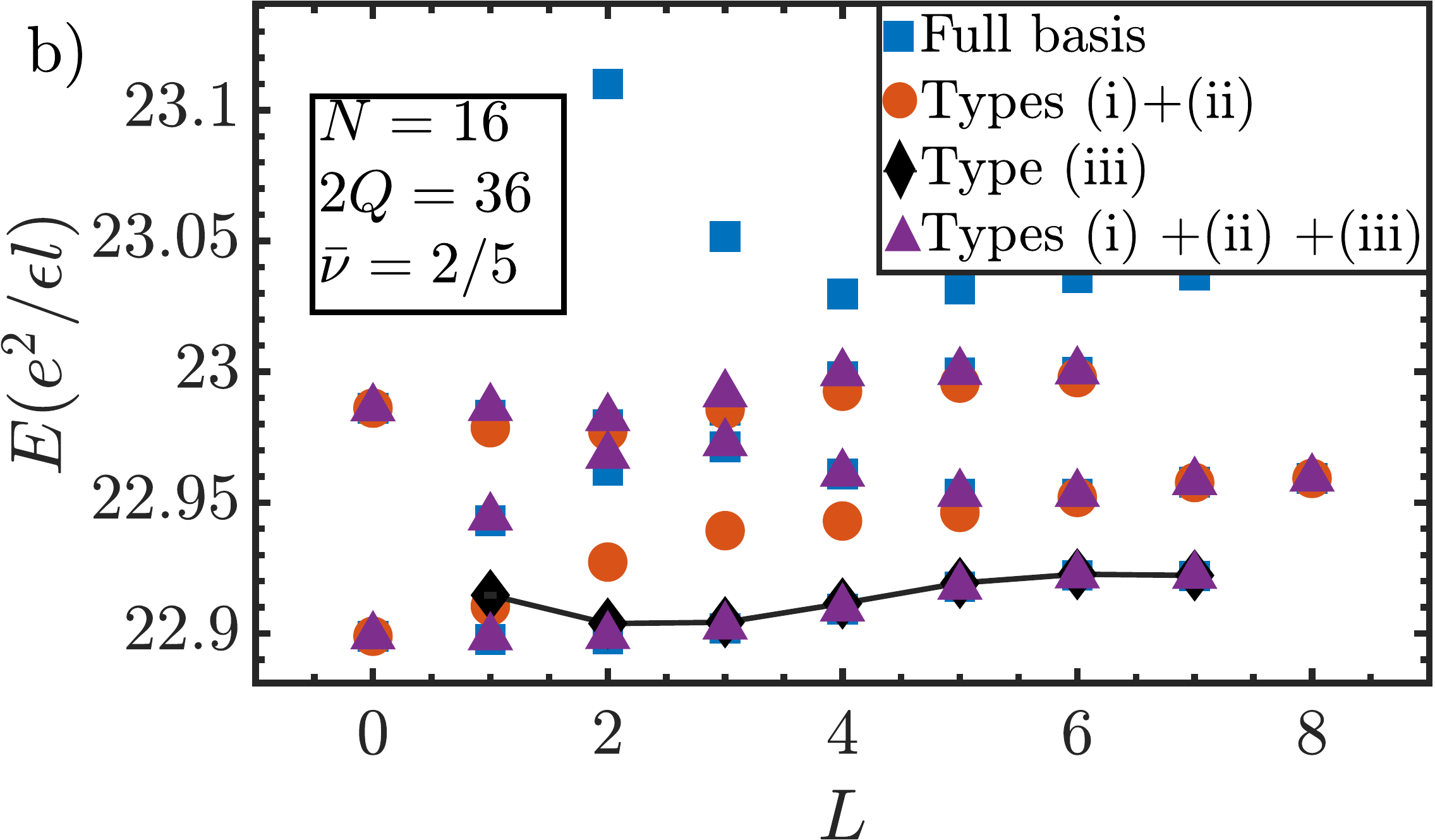}
\caption{(Top) Spectrum of the $\bar\nu = 2/5$ system within the CF basis shown in Fig.~\ref{fig:basis_1_3_2_5}(b). (Bottom) Comparison of energies obtained in various choices of the basis in Fig.~\ref{fig:basis_1_3_2_5}(b), as indicated in the figure. For large $L (\sim k)$, the Type (iii) states give a good approximation to the energy computed in the full basis, while the middle two branches are covered by the type (i) and (ii) excitations. For lower $L$, mixing between the type (iii) and types (i)+(ii) states is needed to bring the energy below the bright state energy.
\label{fig:dispersion_3_5_full}
}
\end{figure}

\begin{figure*}[htb]
\includegraphics[width=0.4\textwidth]{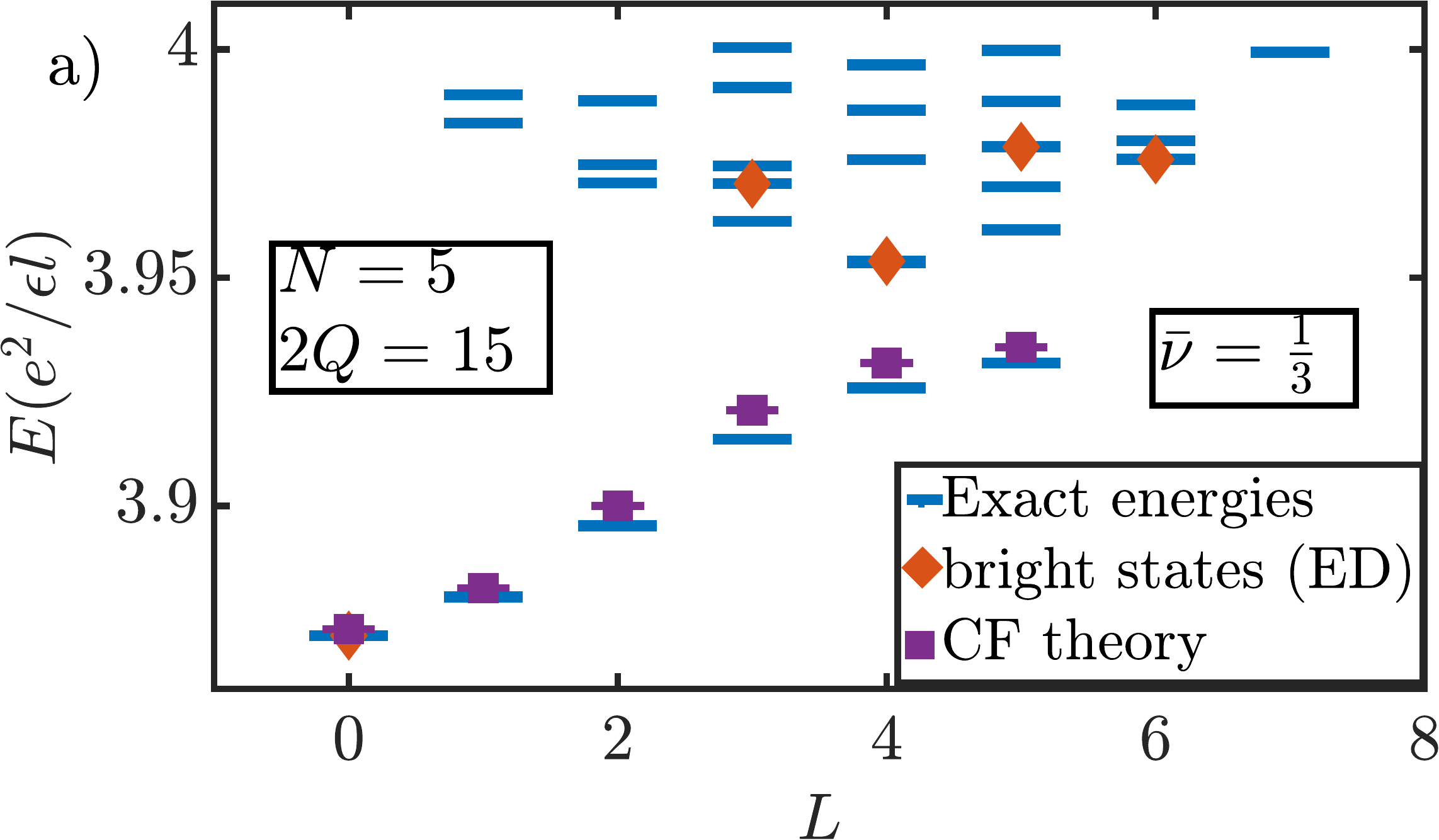}
\includegraphics[width=0.4\textwidth]{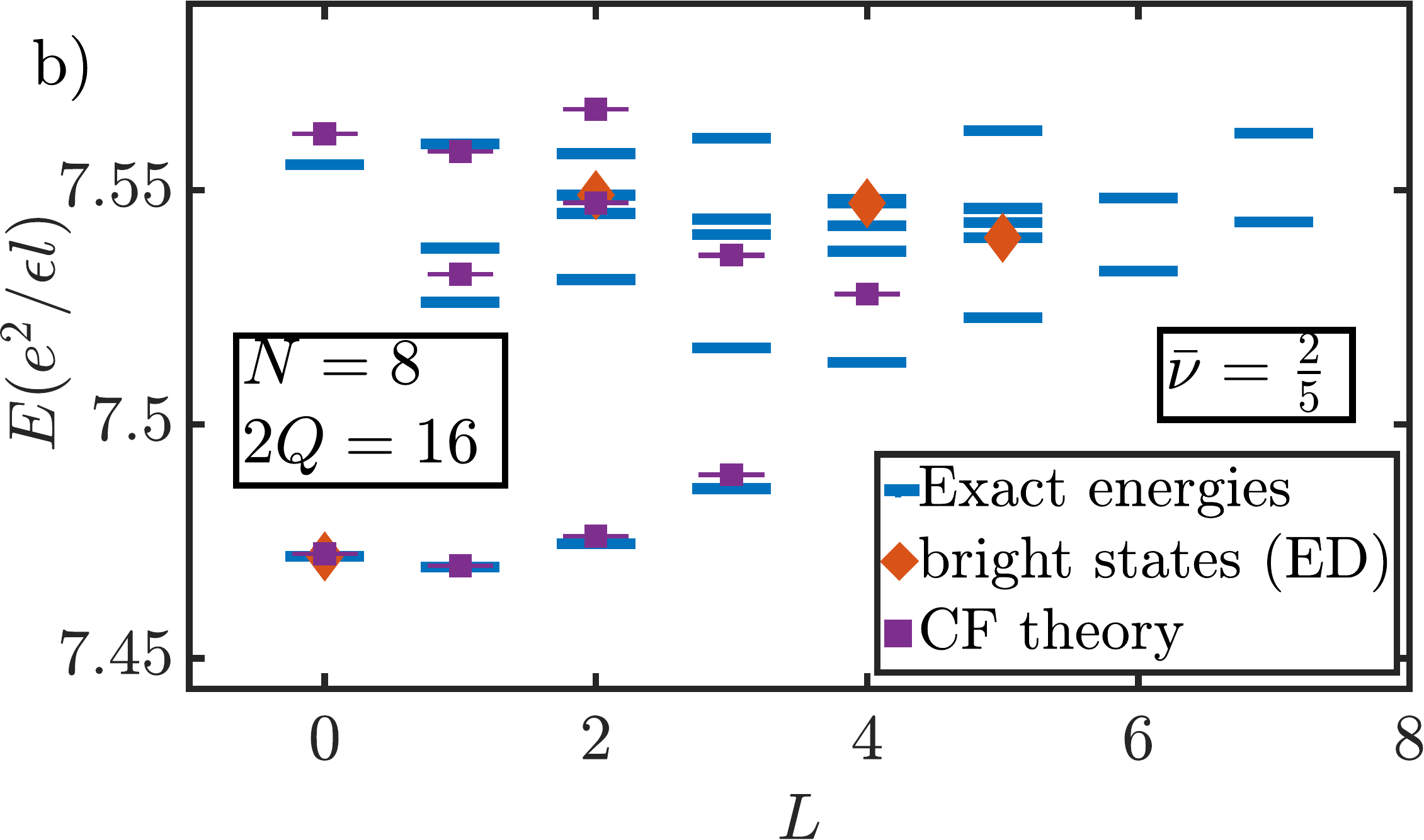}
\includegraphics[width=0.4\textwidth]{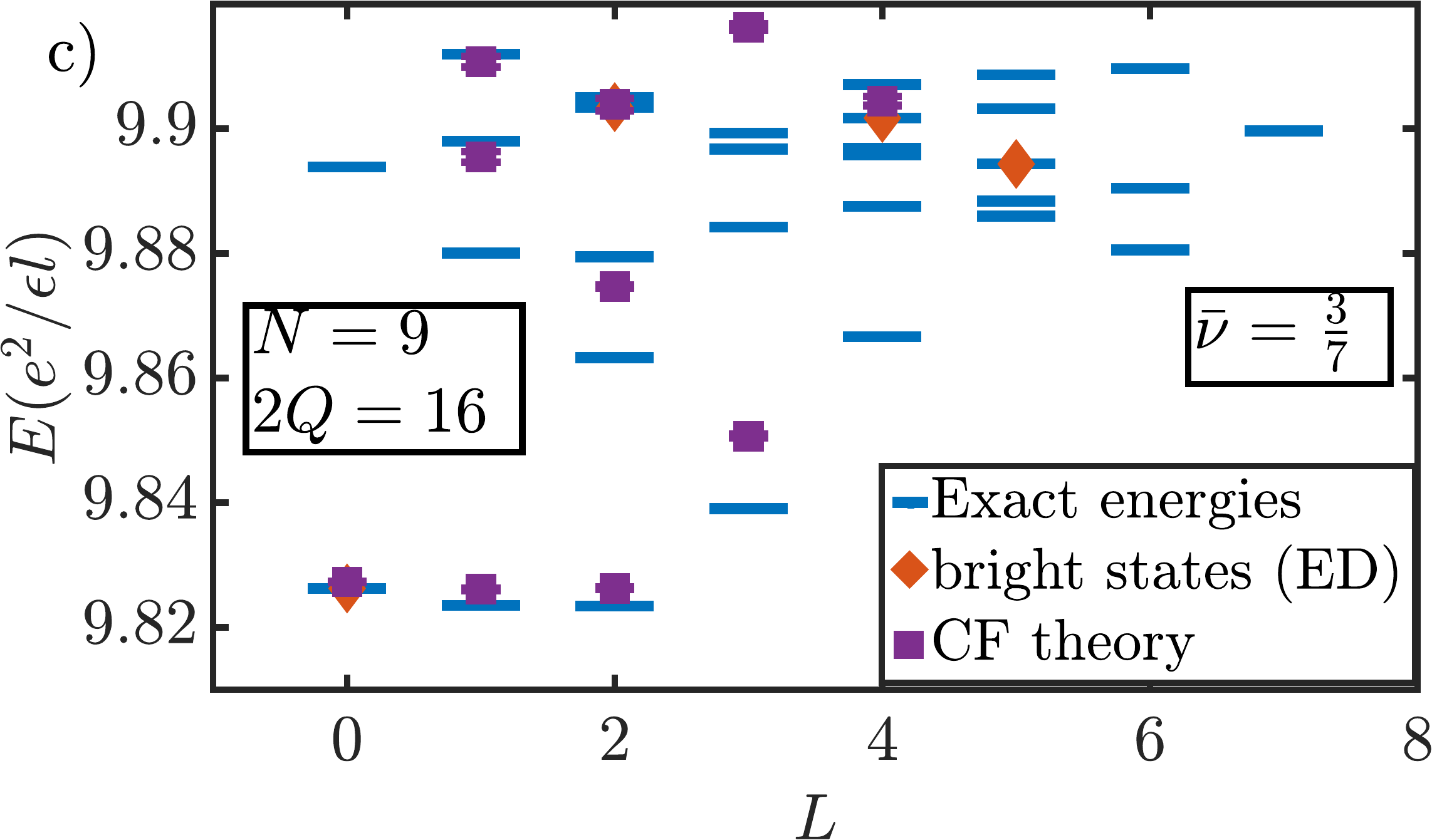}
\includegraphics[width=0.4\textwidth]{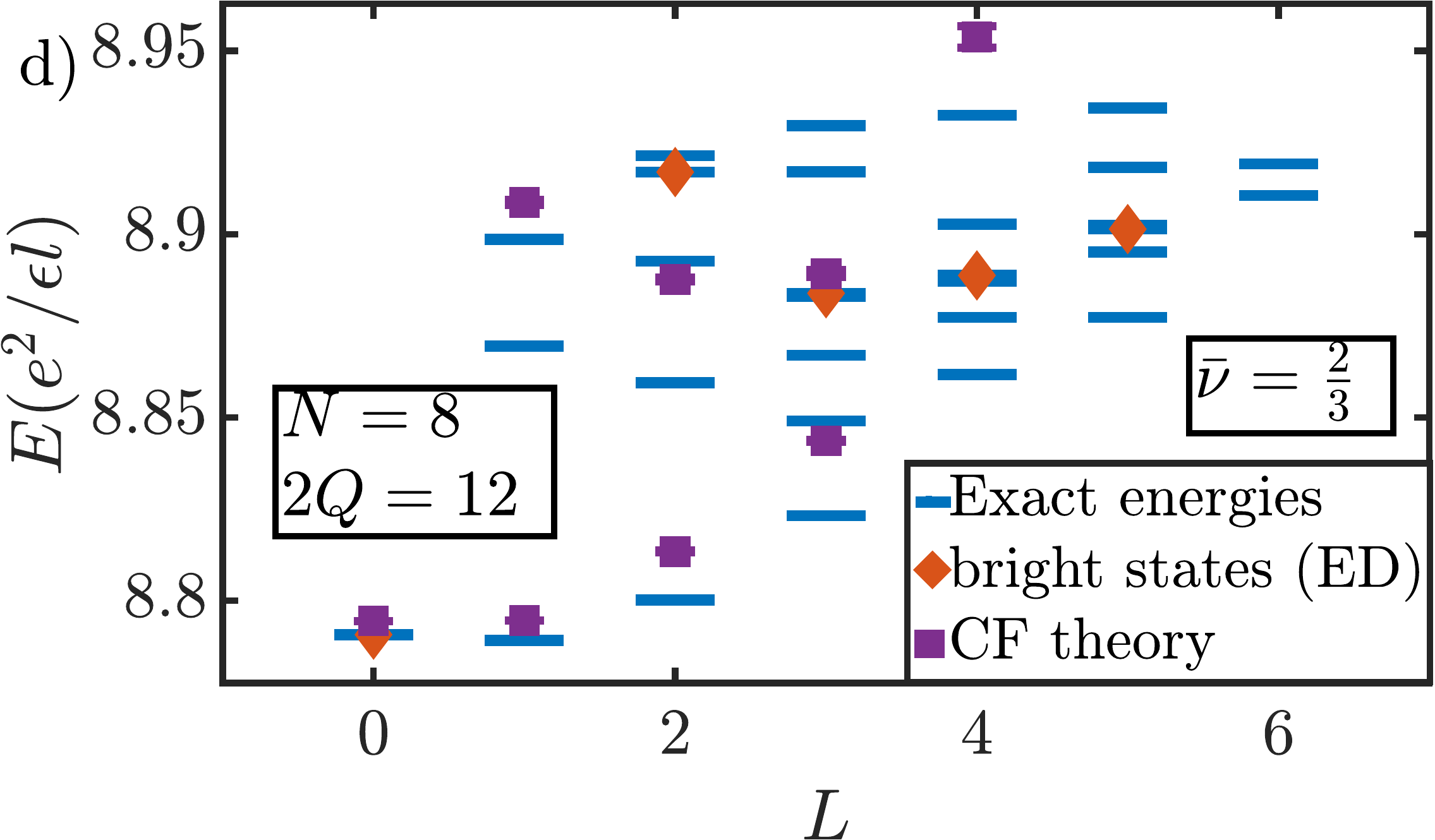}
\caption{Comparison between the small system exact energy spectra and the spectra obtained from CF theory for $\bar\nu = 1/3, 2/5,3/7$ and  $2/3$. CF theory accounts for all the states in the low-energy branch and provides good agreement with the exact values in this range. The ED ground state is dark in all cases except $\bar\nu=1/3$, while the CFD ground state is dark for $\bar\nu = 2/5, 3/7$. At $\bar\nu = 2/3$ 
the lowest bright and dark states from the CFD calculation have the same energies within Monte Carlo error. \label{fig:EDvsCFD_IQL}
}
\end{figure*}

In this section, we first describe the CF bases that we have used in the CF diagonalization (CFD) technique to quantitatively estimate the PL intensity $P(\omega)$ for various fillings $\bar{\nu}$. We next test the validity of our chosen basis by benchmarking against exact diagonalization (ED). We then present results for larger systems, representative of the thermodynamic limit, and show that the important conclusions of the mean-field model described in the main text remain valid even in the presence of interactions.

For Jain fillings $\bar{\nu} = n/(2n+1)$ and $\bar{\nu}=n/(2n-1)$, the basis states consist of a single quasihole in the pseudospin-up \LL s and a single quasiparticle in the pseudospin-down \LL s. We include in our CFD the basis states in which the quasihole lies in the $j\uparrow$\LL\ and the quasiparticle in the $j'\downarrow$ \LL, with $j,j'=0, 1, \cdots n-1$ being the $\Lambda$L index. This leads us to $n^2$ pairs $(j,j')$. We restrict to configurations with $L_z = 0$, since for a particle–hole system with integer total angular momentum $L$, every state has a component with $L_z = 0$. Examples of such bases for $\bar\nu = 1/3$ and $\bar{\nu} = 2/5$ are shown in Fig.~\ref{fig:basis_1_3_2_5}. 

At $\bar\nu = 1/3$, in the non-interacting CF model all the basis states are degenerate. Introduction of interactions leads to the energy spectrum acquiring a dispersion. The dispersion obtained from CFD shows an excellent match with ED as shown in Fig.~\ref{fig:EDvsCFD_IQL}(a). The ground state remains at $L=0$ and is bright. 

At $\bar\nu = \frac{2}{5}$, the CF basis involved is shown in Fig.~\ref {fig:basis_1_3_2_5}(b). In this case, we have more freedom than at $\bar\nu = \frac{1}{3}$ due to the possibility of the quasihole and quasiparticle having different \LL\:  indices. Particularly, there are 4 configurations $(j,j')$ of the quasihole-quasiparticle pair \LL\: indices allowed - i) $(0,0)$, ii) $(1,1)$, iii) $(1,0)$, and iv) $(0,1)$. Measuring the CFKE from the polarized ground state, the states may be classified as following:
\begin{enumerate}
    \item {\bf CFKE conserving states:} The states of type (i) and (ii) have CFKE = $0$ i.e. they have the same (mean-field) energy as the fully polarized ground state. At $\bar\nu = n/(2n+1)$, there are $n$ such states.
    \item {\bf Lower CFKE states:} Type (iii) states have lower CFKEs.
    These states are necessarily dark (see Sec.~\ref{sec:darkness}).
    \item{\bf Higher CFKE states:} These states have higher CFKEs than the fully polarized ground state. At $\bar\nu = 2/5$, such states are shown in (iv). Fig.~\ref{fig:dispersion_3_5_full} shows that the low-energy branch in the spectrum is essentially unaffected by the inclusion of these states in CFD.
\end{enumerate}

\begin{figure}[htb]
\includegraphics[width=0.24\textwidth]{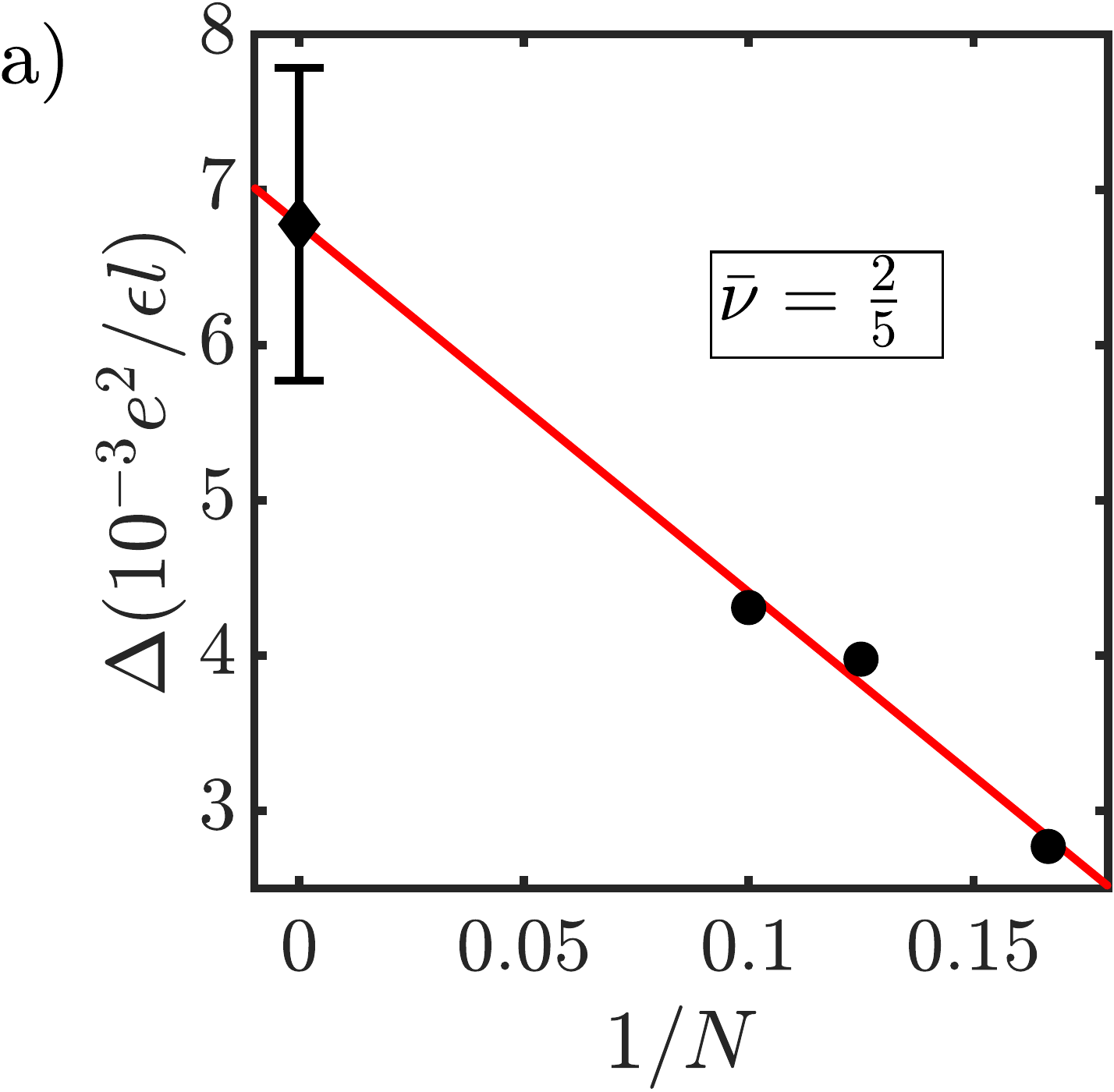}\includegraphics[width=0.24\textwidth]{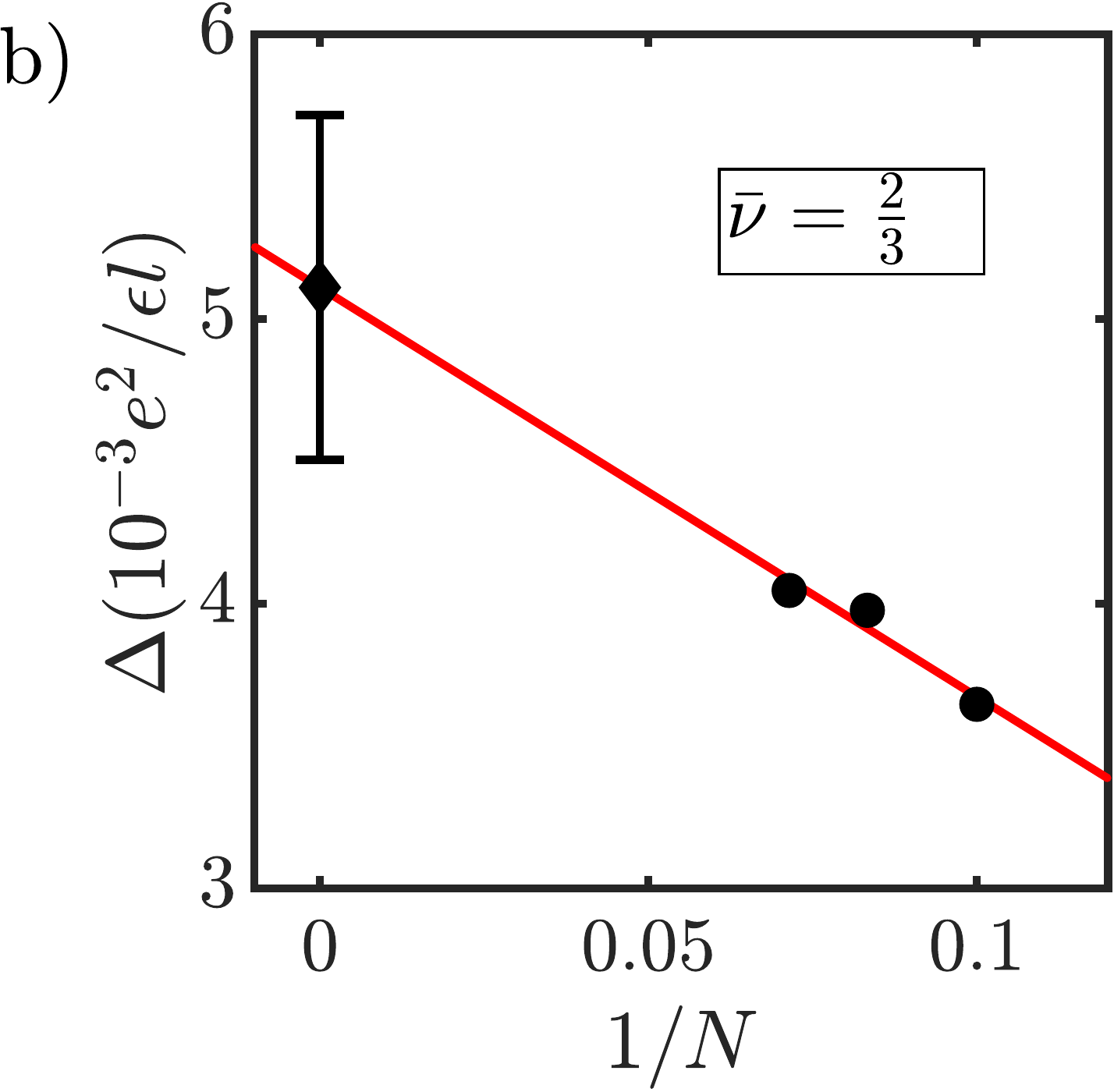}
\caption{Thermodynamic extrapolation for $\Delta$, the energy of the lowest bright state relative to the ground state, for a) $\bar\nu = 2/5$ and b) $\bar\nu = 2/3$. In both cases, the gap survives in the thermodynamic limit with limiting values $\sim 10^{-2}$-$10^{-3} \frac{e^2}{\epsilon l}$. The results are obtained from ED. \label{fig:thermo_ED_2_5_2_3}
}
\end{figure}

The CFD results for the basis described above show excellent agreement with ED, as seen in Fig.~\ref{fig:EDvsCFD_IQL}(b). The ground state is at $L\neq0$ and therefore, for reasons discussed in Sec.~\ref{sec:darkness}, is dark. There are four dispersive pseudospin wave modes (Fig.~\ref{fig:dispersion_3_5_full}(a)) in the spectrum. Fig.~\ref{fig:dispersion_3_5_full}b) shows a comparison between the CF energies obtained from various subsets of the basis described above. The low energy branch is well approximated by the type (iii) states, especially at large $L$. At smaller $L$, some renormalization by mixing with type (i) and type (ii) states is required to get full agreement with the complete CFD basis. Removing type (iv) states from the basis does not change the energies in the low energy branch noticeably.

We work with a similar basis for $\bar\nu =3/7$. The comparison of CFD results with ED is shown in Fig.~\ref{fig:EDvsCFD_IQL}(c), which is again excellent for the low-lying states. The ground state is again at $L\neq 0$ and therefore dark.

For the reverse-flux state at $\bar{\nu} = 2/3$, the agreement with ED is slightly poorer. This is primarily related to how mixed-spin CF wavefunctions are projected into the LLL, as shown in Ref.~\cite{Wu93}. For general mixed-spin CF wavefunctions of the form $\LLL \mathcal{A}[\phi(\Omega_{1}, ..., \Omega_{N}) \phi_{1}^{2}u_{1}u_{2}\dots d_{N}]$, Ref.\cite{Wu93} showed that a “hard-core” projection, in which the LLL wavefunction is constructed as $\mathcal{A}{\phi_{1}\LLL[\phi(\Omega_{1}, ..., \Omega_{N}) \phi_{1}]u_{1}u_{2}\dots d_{N}}$, provides much more accurate wavefunctions than “direct” projection: $\mathcal{A}\{\LLL[\phi(\Omega_{1}, ..., \Omega_{N}) \phi_{1}^{2}]u_{1}u_{2}\dots d_{N}\}$. However, since no local projection approach like the Jain-Kamilla method is known for hard-core projection (and we are thus forced to resort to an expansion in terms of basis functions), its complexity scales super-exponentially with yhe system size, making it difficult to work with in practice. We have thus obtained CF results only on the basis of “direct” projection implemented by the Jain-Kamilla method, which is much easier to work with. Because the agreement between the CFD and ED is not sufficiently accurate here, we rely on ED to estimate the thermodynamic behavior at $\bar{\nu} = 2/3$.

\begin{figure*}[t]
\includegraphics[width=0.24\textwidth]{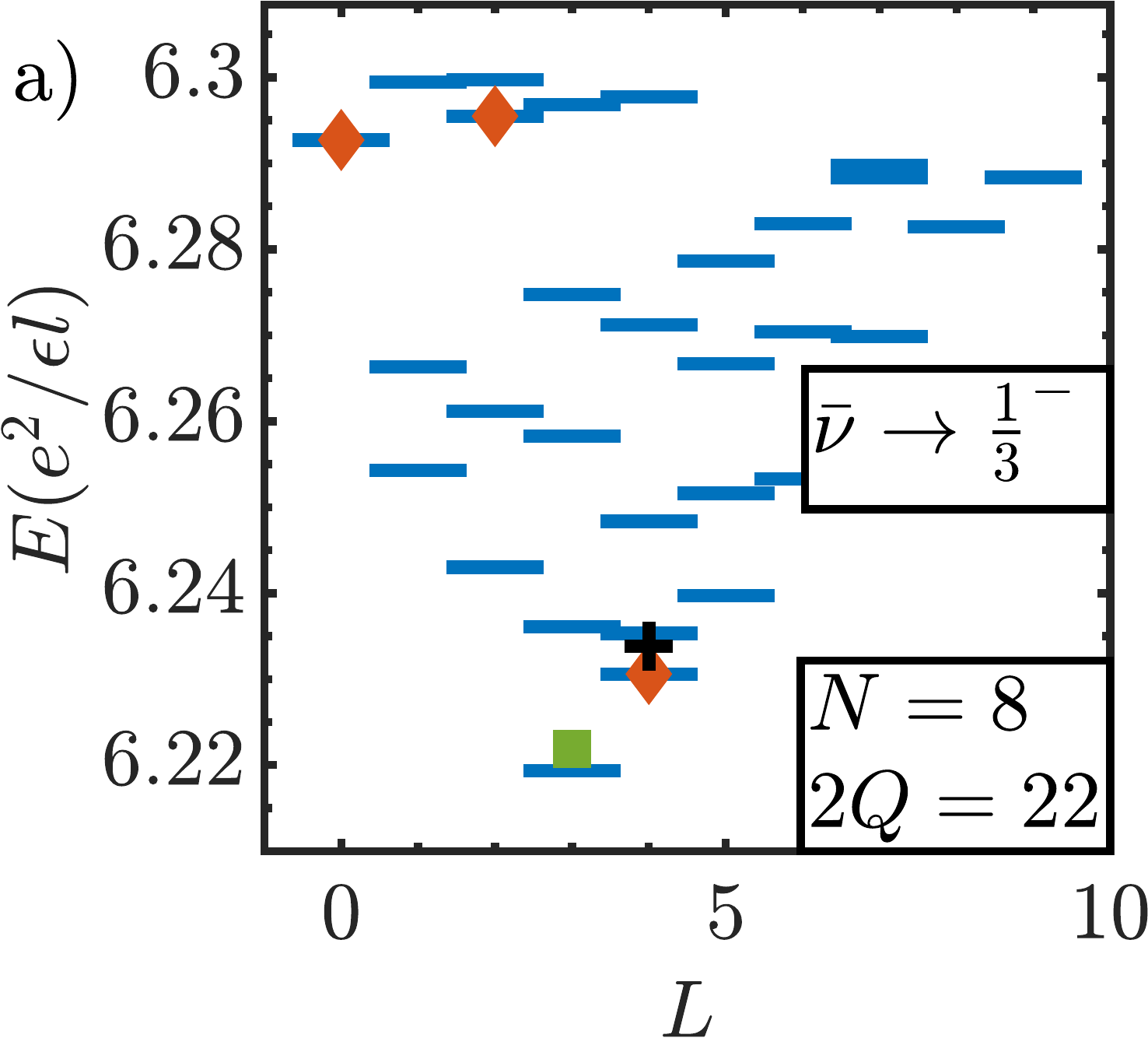}\includegraphics[width=0.24\textwidth]{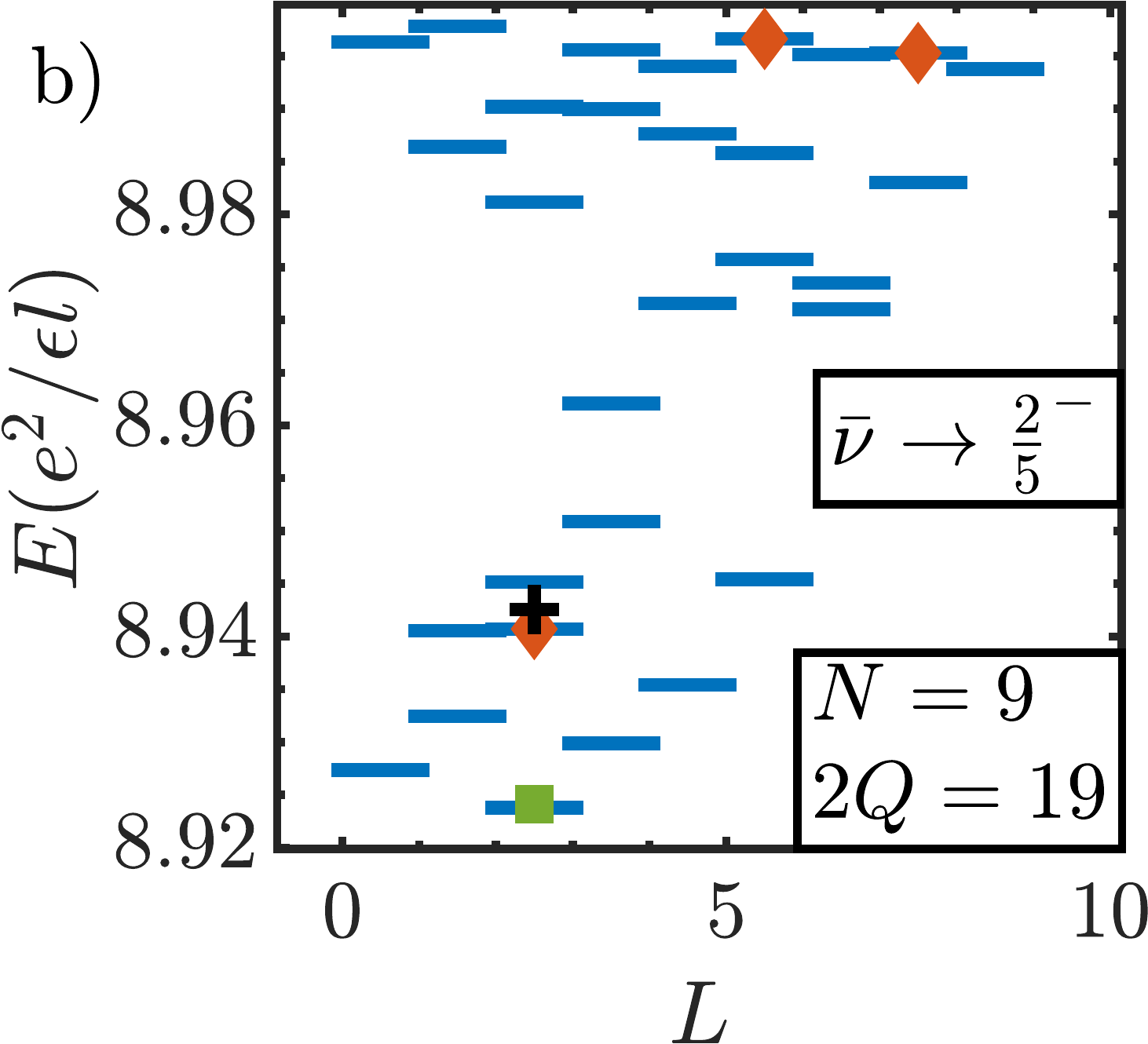}\includegraphics[width=0.24\textwidth]{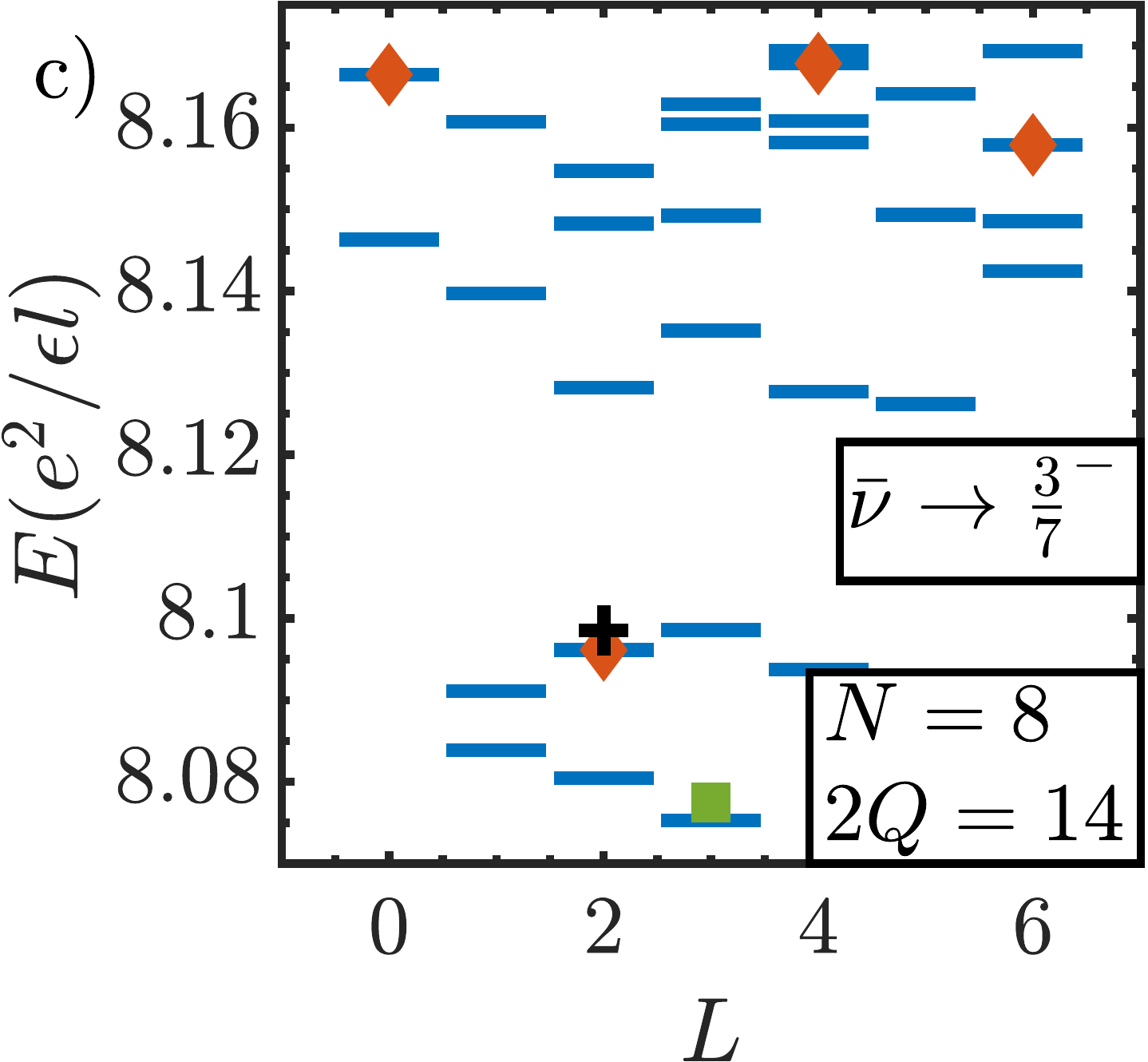}\includegraphics[width=0.247\textwidth]{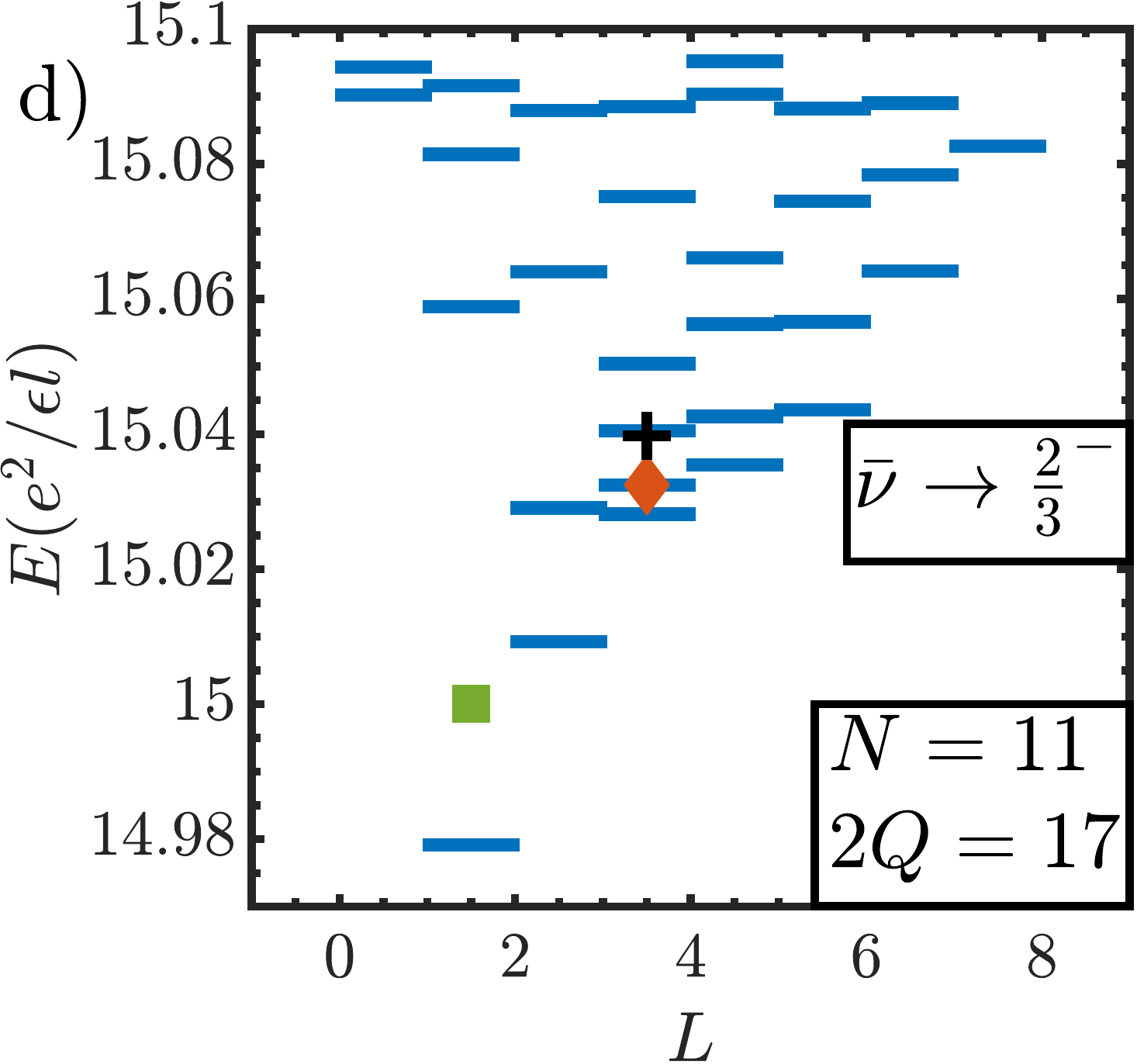}
\includegraphics[width=0.236\textwidth]{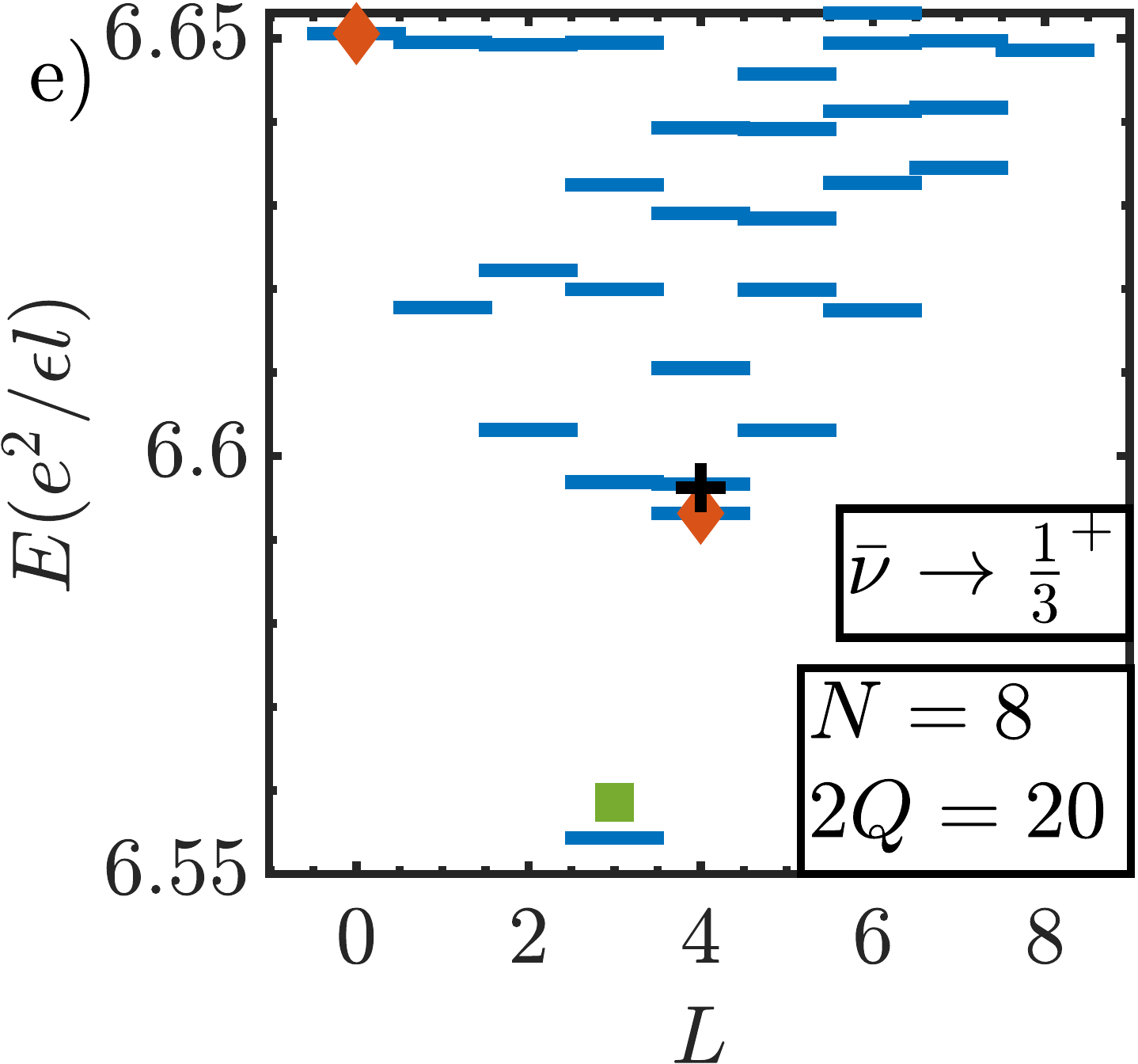}\includegraphics[width=0.236\textwidth]{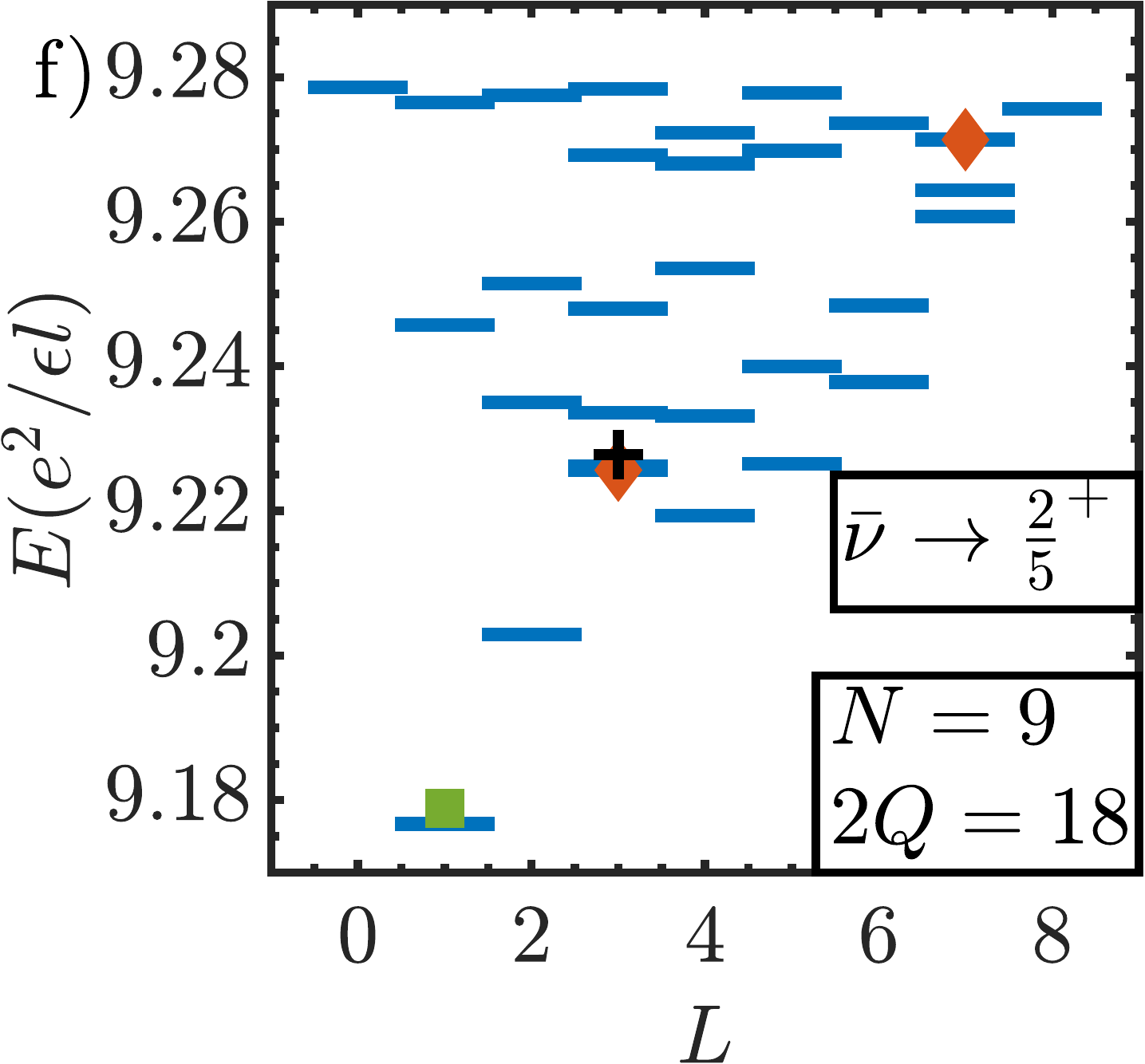}\includegraphics[width=0.245\textwidth]{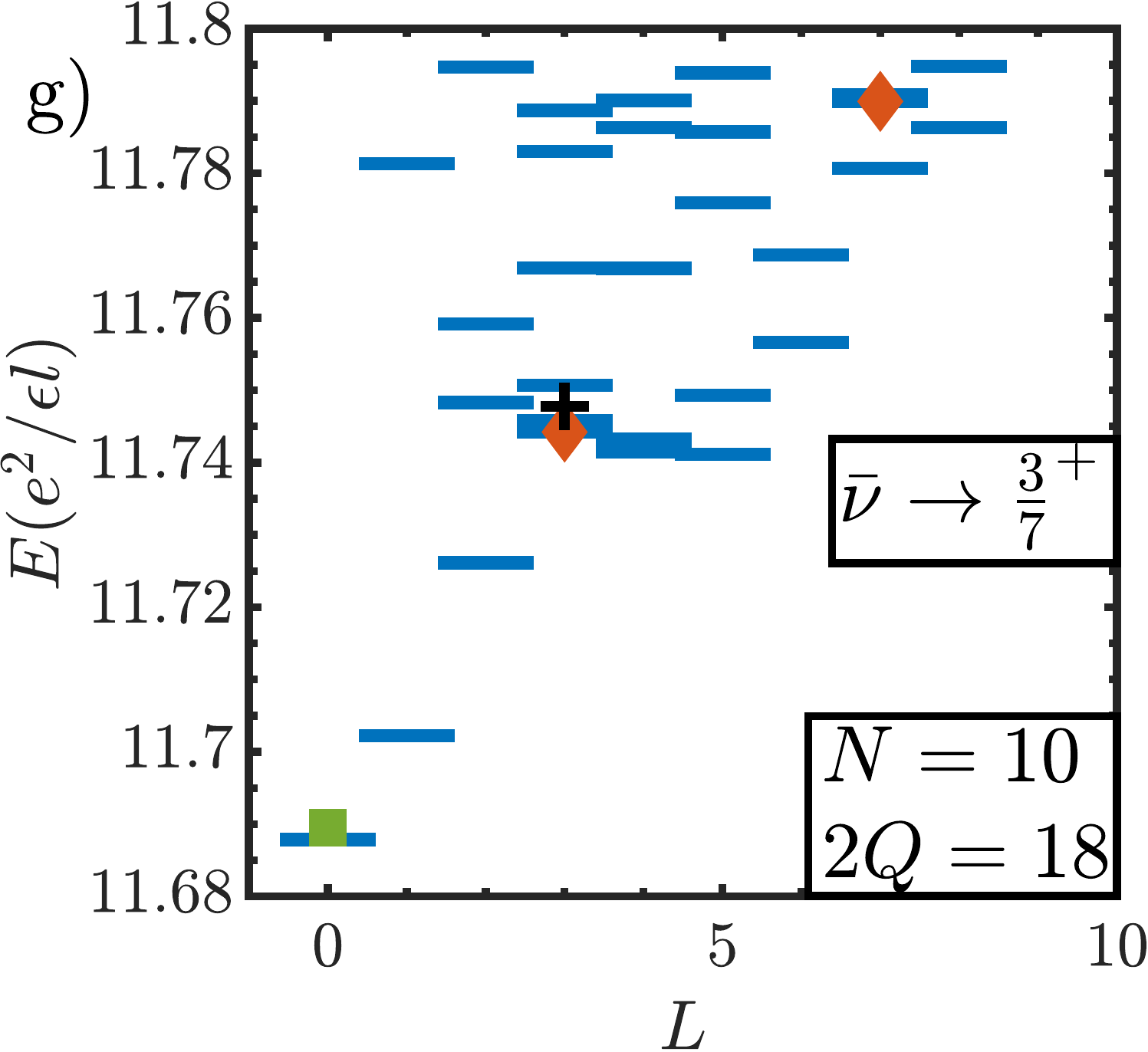}\includegraphics[width=0.245\textwidth]{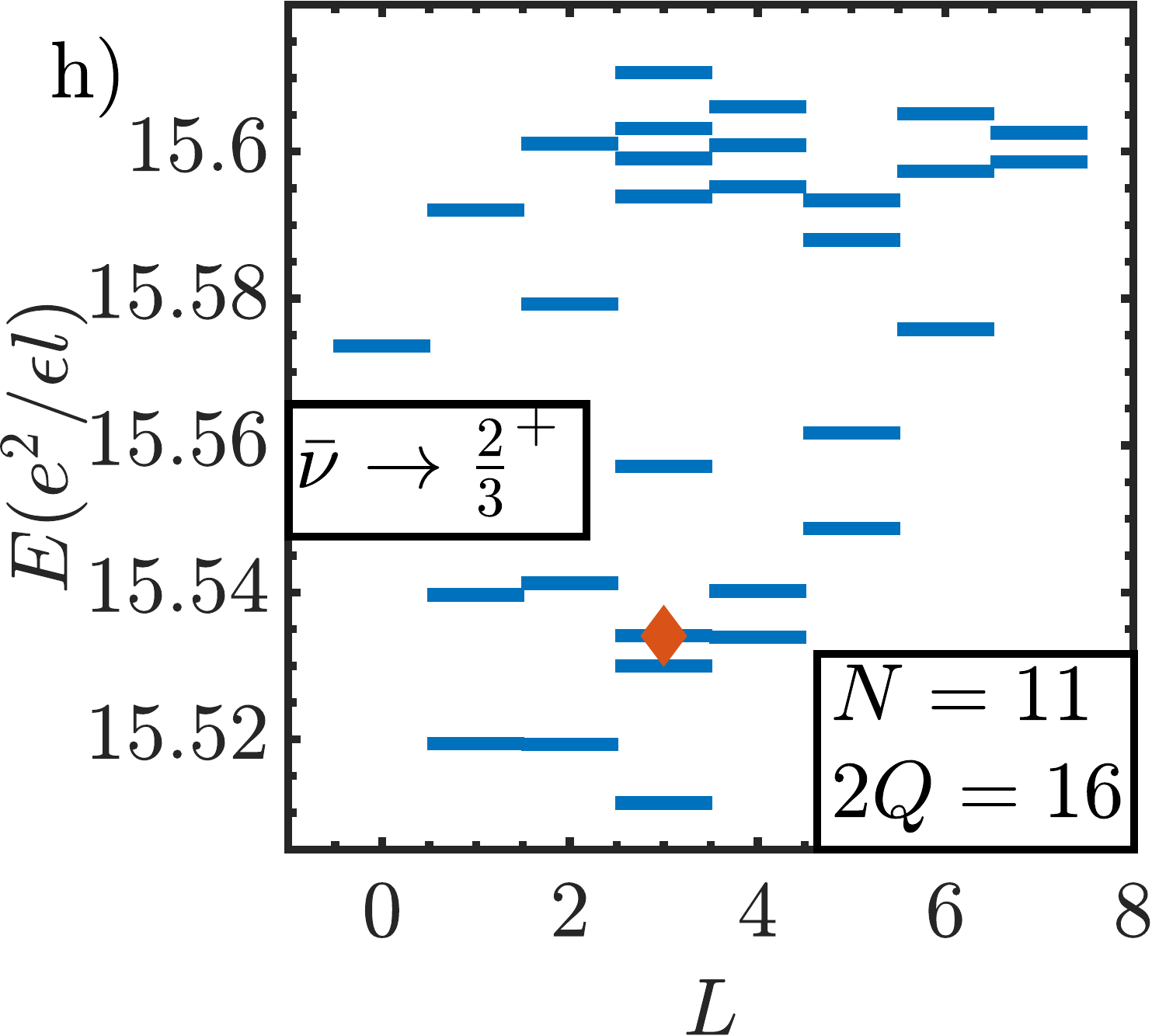}
\includegraphics[width=0.55\textwidth]{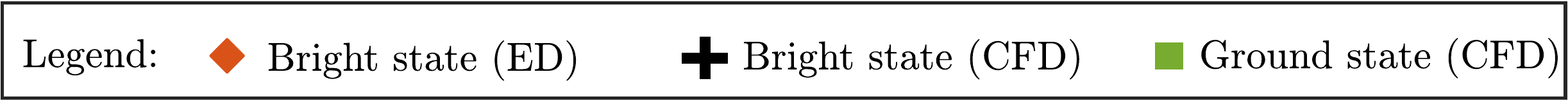}
\caption{ (a-d)Exact energy spectra for single quasihole states near $\bar\nu = \frac{1}{3},\frac{2}{5},\frac{3}{7},\frac{2}{3}$. The ground state is dark in all the cases. (e-h)  Spectra for single quasiparticle states near $\bar\nu = \frac{1}{3},\frac{2}{5},\frac{3}{7},\frac{2}{3}$. The ground state is again dark in all the cases. (a-g) also show a comparison with CFD values of the lowest bright state (black ``+'') and the ground state energies (green square), computed within the bases described in the text. At all parallel-flux states, the agreement between CFD and ED is excellent.
\label{fig:ED_QH_QP}
}
\end{figure*}

At $\bar{\nu} = 2/3$, the energy difference between the Jain (lowest-energy bright) state and the $S_{z} = N/2 - 1$ ground state (which is dark), shown in Fig.~\ref{fig:thermo_ED_2_5_2_3}, is $\sim 0.005e^{2}/\varepsilon\ell$, consistent with the picture in Sec.~\ref{sec:darkness}. For $\bar{\nu} = 2/5$, an ED-based extrapolation yields a corresponding energy gap of $\sim 0.007e^{2}/\varepsilon\ell$, several times larger than the CFD value of $\sim 0.0012e^{2}/\varepsilon\ell$, although the qualitative picture remains intact.

We now turn to states away from the Jain fillings $\bar{\nu} = n/(2n+1)$, beginning with systems containing a single quasiparticle (QP), i.e., $\bar{\nu} \gtrsim n/(2n+1)$. As discussed previously, within the non-interacting CF picture, the $S_{z} = N/2 - 1$ ground state corresponds to a single down-spin CF occupying the lowest $\Lambda$ level, while the remaining up-spin CFs fill the first $n$ $\Lambda$Ls completely (see for e.g. Fig.~\ref{fig:Schematic3}(c)). This configuration has total angular momentum $L = Q^{\star}$. A comparison with ED results (Fig.~\ref{fig:ED_QH_QP}) shows that the lowest-energy state indeed occurs at $L = Q^{\star}$ and matches the energy of the corresponding CF state very closely. This confirms that, for $\bar{\nu} \gtrsim n/(2n+1)$, the $S_{z} = N/2 - 1$ ground state cannot be bright. This can be understood by considering the lowest-energy $S_{z} = N/2$ state, where a single up-spin CF occupies the $n^{\rm th}$ $\Lambda$L, with the lower $n$ $\Lambda$Ls fully filled. This state has angular momentum $L = Q^{\star} + n$. The corresponding bright state, obtained by applying $S^{-}$, also has $L = Q^{\star} + n$, which is never equal to $Q^{\star}$ for any $n$.

For states with $\bar{\nu} \lesssim n/(2n+1)$—specifically, systems containing two up-spin CF holes in the first $n$ $\Lambda$Ls and one down-spin CF (see, e.g., Fig.~\ref{fig:Schematic3}(e))—the non-interacting CF picture predicts that the lowest-energy configuration has the two holes in the $(n-1)^{\rm th}$ $\Lambda$L and the down-spin CF in the $0^{\rm th}$ $\Lambda$L. Treating the residual interaction among these three as that of point particles, we expect them to cluster and form a bound trion. Due to the Pauli principle, the minimum relative angular momentum of the two holes is 1, giving a pair angular momentum of $2(Q^{\star}+n-1) - 1 = 2Q^{\star} + 2n - 3$. When the down-spin CF ($L = Q^{\star}$) is brought as close as possible to this pair, the resulting trion has angular momentum $L = Q^{\star} + 2n - 3$. We thus predict that the lowest-energy $S_{z} = N/2 - 1$ state at $\bar{\nu} \lesssim n/(2n+1)$ is a trion at $L = Q^{\star} + 2n - 3$.

\begin{figure}[htb]
\includegraphics[width=0.45\textwidth]{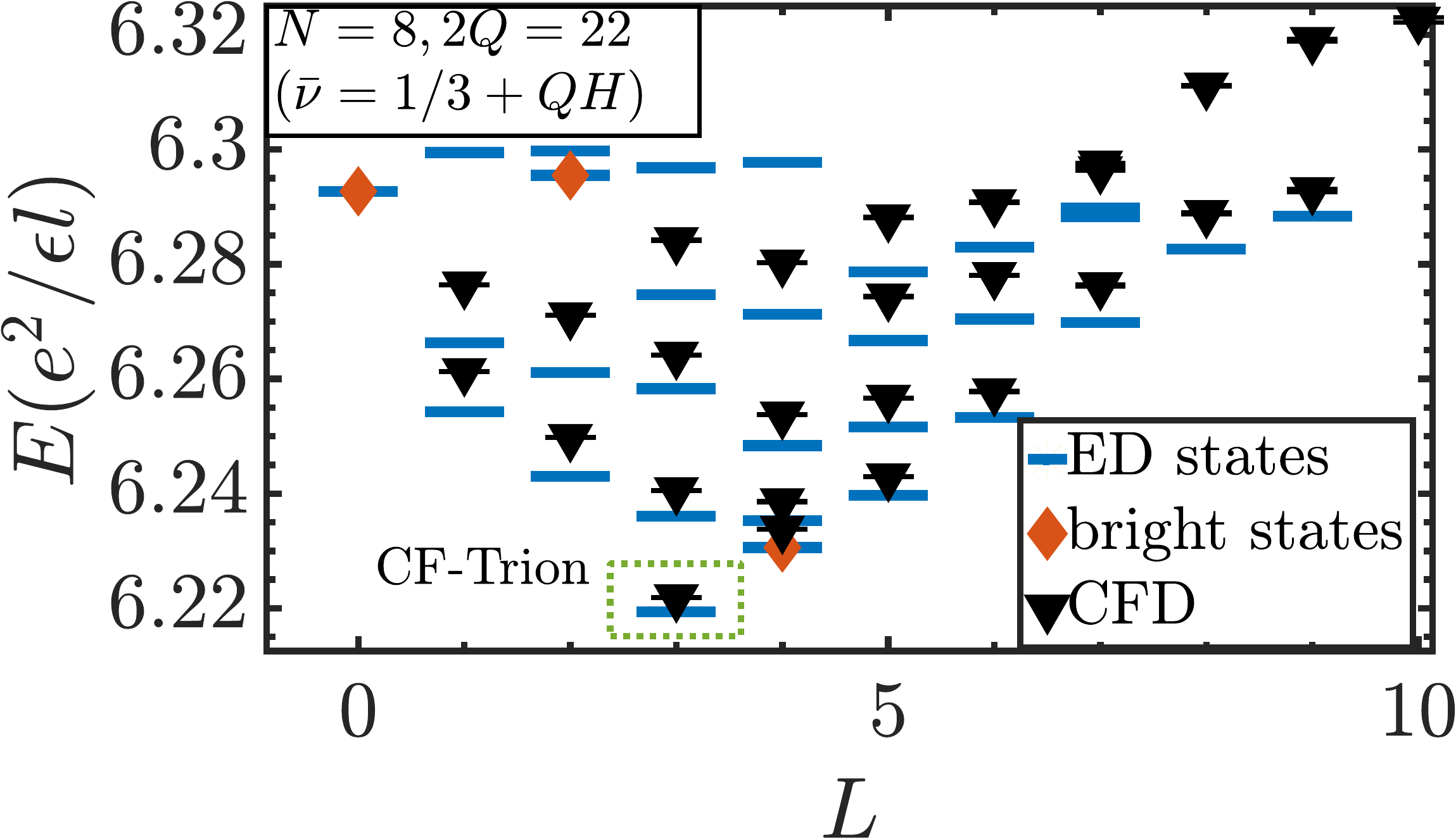}
\caption{CFD and ED spectra for the quasihole state near $\bar\nu = 1/3$ for a small system. The ground state is dark and is separated from the lowest bright state by a gap of the order $10^{-2} \frac{e^2}{\epsilon l}$. The ground state is a CF analog of the previously known dark-trion bound state of electrons\cite{Wojs95}.\label{fig:QH_1_3_CFD_vs_ED}
}
\end{figure}

For $\bar{\nu} \lesssim 1/3$, we perform CFD in the $S_{z} = N/2 - 1$, $L_{z} = 0$ basis consisting of two up-spin CF holes and one down-spin CF. As shown in Fig.~\ref{fig:QH_1_3_CFD_vs_ED}, the ground state appears at $L = Q^{\star} + 2n - 3 = Q^{\star} - 1$, confirming the trion picture. This state has a different angular momentum than the lowest-energy bright state, which is obtained by applying $S^{-}$ to a state with a single down-spin CF hole in the $0^{\rm th}$ $\Lambda$L (at $L = Q^{\star}$), and is thus dark.

For $\bar{\nu} \lesssim n/(2n+1)$ with $n = 2, 3, \dots$, we perform CFD in a restricted basis at $S_{z} = N/2 - 1$, $L_{z} = 2Q^{\star} + 2n - 3$. By limiting to states with $L \geq 2Q^{\star} + 2n - 3$, we reduce the basis size thus reducing computing cost and enabling extrapolation to the thermodynamic limit. In all cases, the ground state occurs at $L = 2Q^{\star} + 2n - 3$, consistent with ED results (Fig.~\ref{fig:ED_QH_QP}) and confirming the formation of a trion. Further, a comparison with the energy of the lowest bright states, also establishes that the trion is dark i.e. is an $S = N/2-1$ state.

For states slightly away from the reverse-flux fillings $\bar{\nu} = n/(2n-1)$—i.e., at $\bar{\nu} \lesssim n/(2n-1)$ and $\bar{\nu} \gtrsim n/(2n-1)$—the situation is effectively the reverse of that near the parallel-flux fillings $\bar{\nu} = n/(2n+1)$. Specifically, the behavior at $\bar{\nu} \lesssim n/(2n-1)$ mirrors that at $\bar{\nu} \gtrsim n/(2n+1)$, and vice versa. Accordingly, the angular momentum rules for the ground state are exchanged between these cases. This is confirmed by ED results at $\bar{\nu} \gtrsim 2/3$ shown in Fig.~\ref{fig:ED_QH_QP}(d), and by both ED and CFD results shown at $\bar{\nu} \lesssim 2/3$ shown in Fig.~\ref{fig:ED_QH_QP}(h). As with the incompressible state at $\bar{\nu} = 2/3$ discussed earlier, the CFD energies obtained from JK projection in this regime are less accurate than for $\bar{\nu} < 1/2$.

\begin{figure*}
\includegraphics[width=0.40\textwidth]{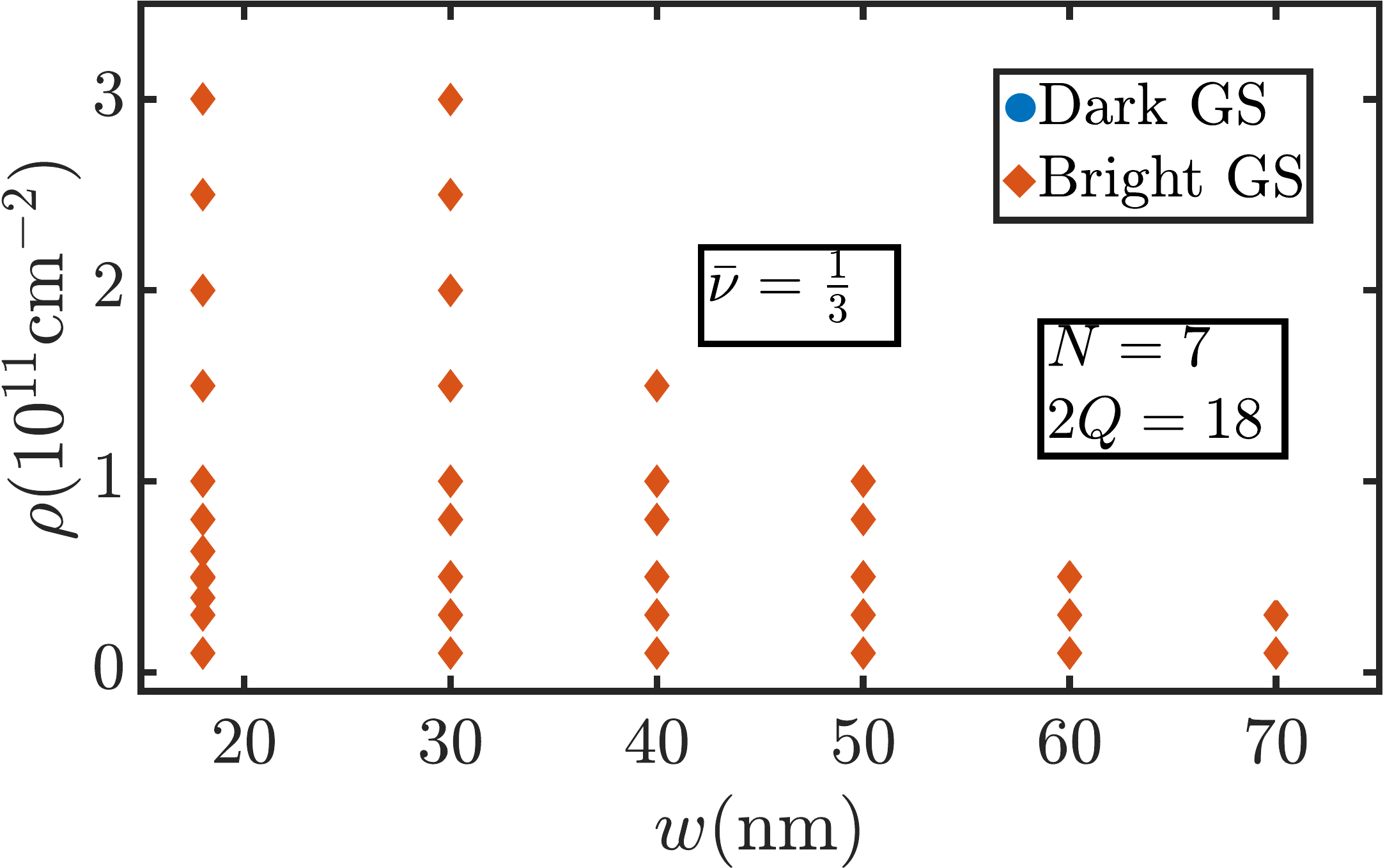}\hspace{1cm}\includegraphics[width=0.40\textwidth]{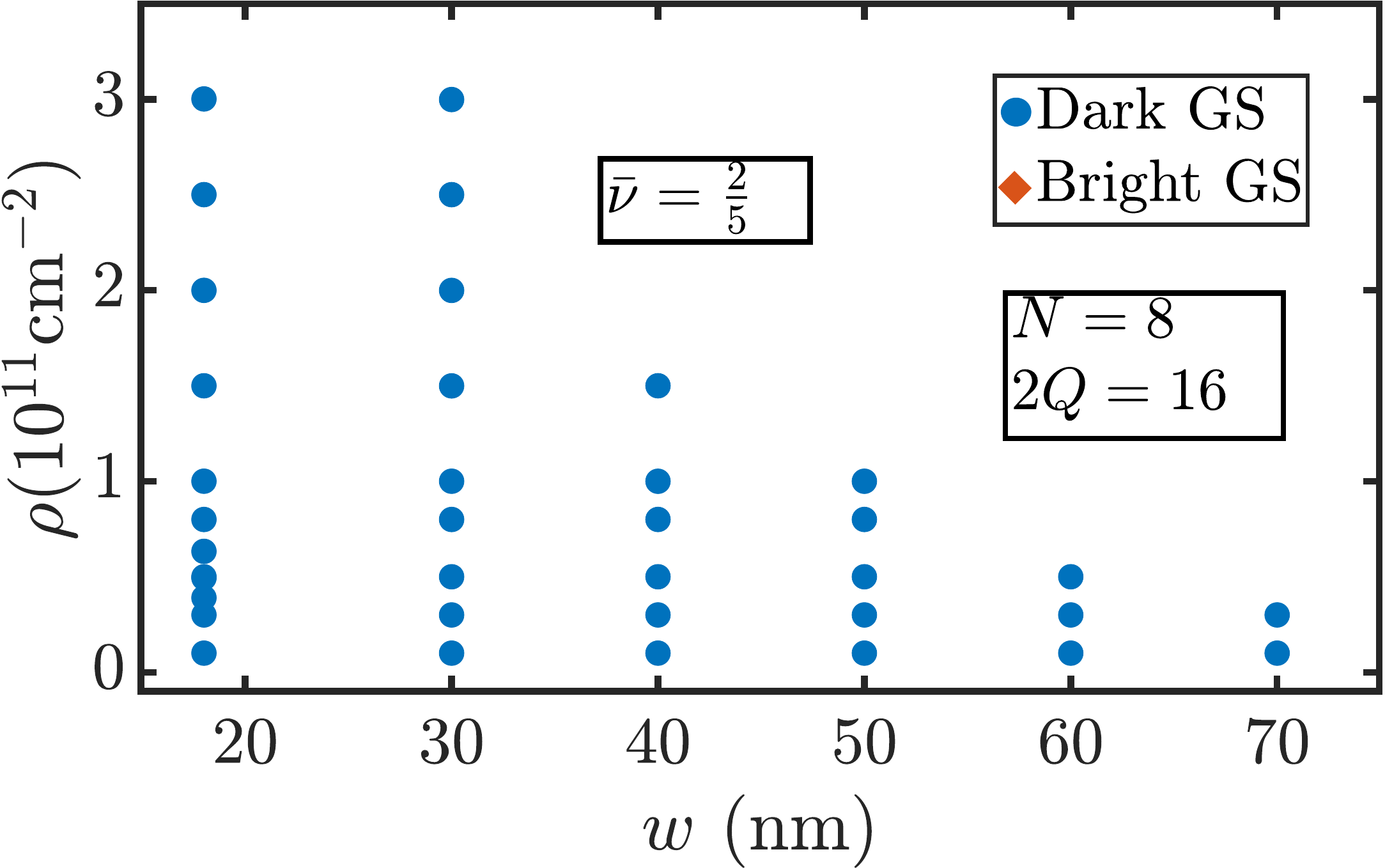}\vspace{1cm}
\includegraphics[width=0.40\textwidth]{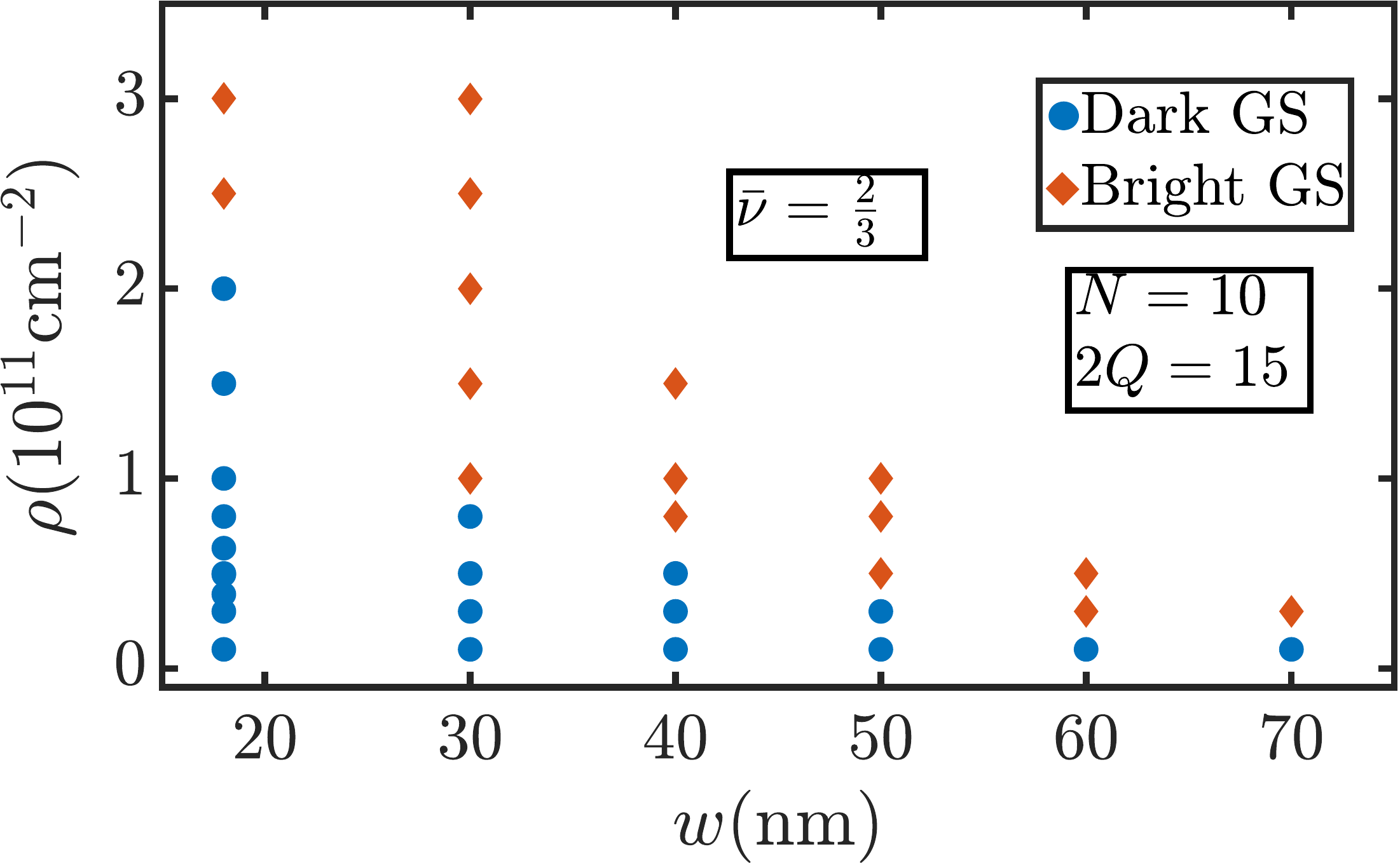}\hspace{1cm}\includegraphics[width=0.40\textwidth]{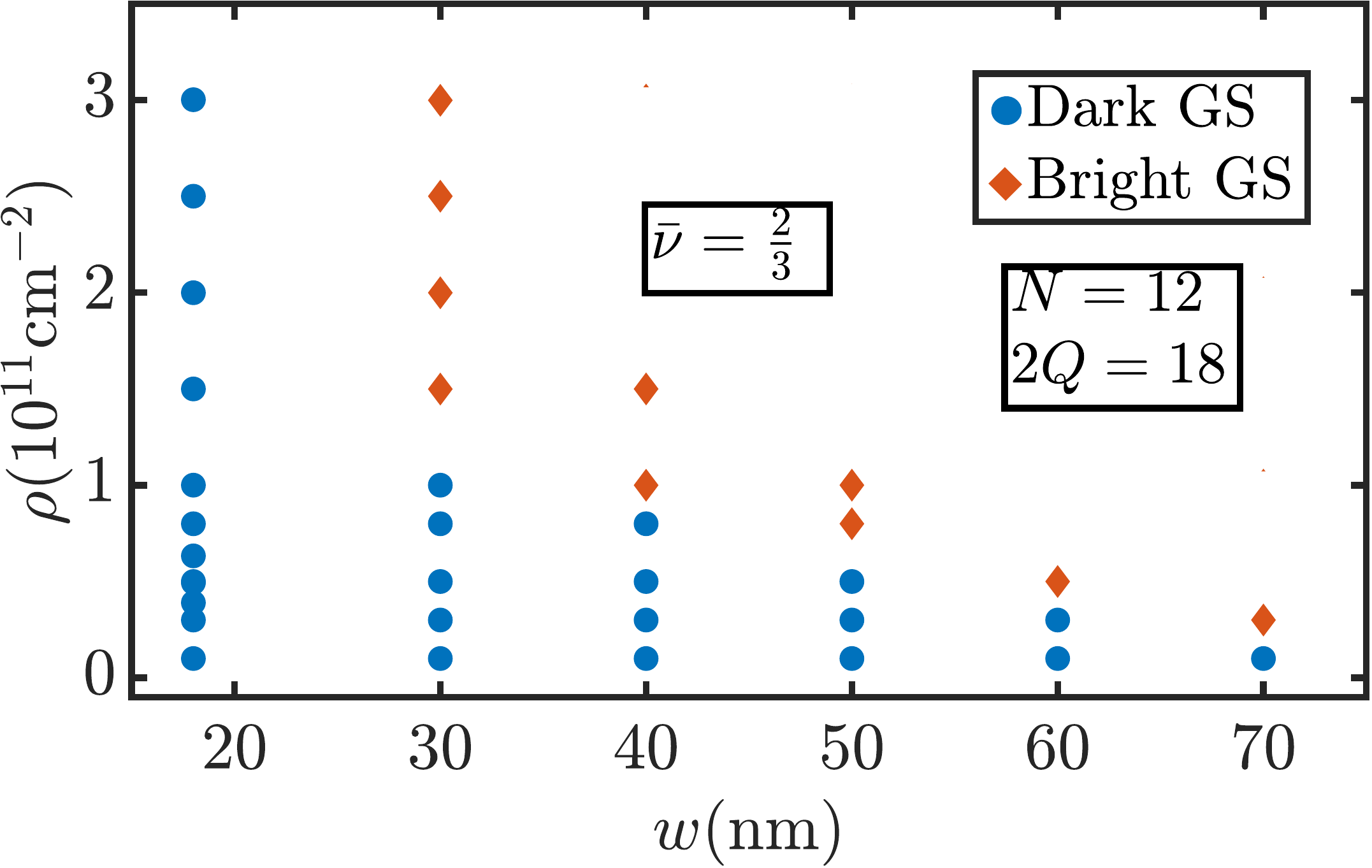}
\caption{
Top Left: Phase diagram of the $\bar\nu = 1/3$ state. The ground state in all cases is found to be bright. Top Right: Phase diagram of the $\bar\nu = 2/5$ state. The ground state in all cases is found to be dark.
Bottom: Phase diagrams for $\bar\nu = 2/3$ for two different system sizes. There is a ``bright" phase at high well widths/high densities, where the ground state becomes bright. In the other phase the ground state is dark. This phase contains the infinitely thin well limit. The phase boundary changes with system size implying that the thermodynamic limit is not reached at these system sizes. \label{fig:phase_diagram_finite_width} }
\end{figure*}

\section{Can finite width brighten a FQHE state?}
\label{sec:brighten}

In this section, we investigate the effects of quantum well width using ED, while continuing to assume $SU(2)$ symmetry—that is, electrons and holes are taken to have identical transverse wavefunctions. Our focus is on whether a dark state at $T = 0$ can be rendered bright—or vice versa—by tuning the quantum well width $w$ or electron density $\rho$. We find that while most states do not change, the $\bar\nu=2/3$ state can become bright with increasing quantum-well width.

We consider a two-dimensional electron gas confined in a GaAs quantum well, with transverse wavefunctions modeled using the LDA approximation~\cite{Ortalano97,Zhang86}. The systems studied span well widths $w = 18$–$70$ nm and electron densities $\rho = (1$–$30) \times 10^{10}\mathrm{cm}^{-2}$. As shown in Fig.~\ref{fig:phase_diagram_finite_width}, at $\bar{\nu} = 1/3$ and $\bar{\nu} = 2/5$, the results remain unchanged from the ideal case: the $\bar{\nu} = 1/3$ state is bright, while the $\bar{\nu} = 2/5$ state is dark.

The $\bar{\nu} = 2/3$ case is more interesting, as shown in Fig.~\ref{fig:phase_diagram_finite_width}. In the ideal limit, as discussed earlier, the $\bar{\nu} = 2/3$ state produces no PL signal at $T=0$. However, we find that softening the short-range part of the interaction—by increasing $w$ or $\rho$—can lead to a bright $S_{z} = N/2 - 1$ ground state, i.e., an $S = N/2$ state. In the system sizes accessible to ED, we are unable to reliably determine the precise “phase boundary” at which the $\bar{\nu} = 2/3$ state transitions between dark and bright. However, we do expect, the general trend to survive in the thermodynamic limit i.e. softening the short-range part of the interaction will lead to a bright $\bar{\nu} = 2/3$ state.

\section{Results with SU(2) breaking perturbations}\label{app:su2break}

\begin{figure}[htb]
\includegraphics[width=0.4\textwidth]{ED_spectrum_80_pct_8_21_-6_0.pdf}
\includegraphics[width=0.4\textwidth]{PL_spectrum_80_pct_8_21_-6_0.pdf}
\caption{{Top: ED spectrum at $\bar\nu = 1/3$ for $\kappa = 0.8$. The ground state is adiabatically connected to the bright ground state of the $SU(2)$ symmetric system.
Bottom: PL spectrum for the same system. Since the ground state is nearly bright, a peak is visible even at $T = 0$. The peak comes from the transition shown by the green arrow in the top figure.} 
\label{fig:80_pct_1_3}
}
\end{figure}

\begin{figure}[htb]
\includegraphics[width=0.4\textwidth]{ED_spectrum_80_pct_9_18_-7_0.pdf}
\includegraphics[width=0.4\textwidth]{PL_spectrum_80_pct_9_18_-7_0.pdf}
\caption{{Top: ED spectrum for 1 QP at $\bar\nu = 2/5$ for $\kappa = 0.8$. The trion energy is $\Delta = 0.0394$.
Bottom: PL spectrum for the same system. At $T\ll \Delta$, the total intensity is very small, but as $T$ becomes of the same order of magnitude as $\Delta$, we see the appearance of large peak originating from the transition between the nearly bright state and the ground state of the $S_z = N/2$ sector (green arrow in top figure).} 
\label{fig:80_pct_2_5_qp}
}
\end{figure}

Fig.~\ref{fig:80_pct_1_3} shows the energies and the PL spectrum for a system with $\kappa = 0.8$ at $\bar\nu = 1/3$. We see a PL peak at $T = 0$, as in the $SU(2)$ symmetric case. This peak remains the most prominent upto at least $T = 0.015$, at which point the Boltzmann population of the higher energy states starts to become significant and we start to see the emergence of side peaks.

The typical case of PL at a compressible filling fraction is illustrated in Fig.~\ref{fig:80_pct_2_5_qp} for 1 QP of $\bar\nu = 2/5$. At temperatures much less than the gap $\Delta$ between the ground state and the lowest "nearly bright" state (state connected adiabatically to the lowest bright state), the total PL intensity is extremely small. As the temperature becomes of the order of $\Delta$, we see the emergence of a strong peak corresponding to the transition between the lowest nearly bright state to the ground state of the $S_z = N/2$ system. Weak side peaks are also present, arising from various other transitions. 

\begin{figure}[htb]
\includegraphics[width=0.4\textwidth]{ED_spectrum_50_pct_8_16_-6_0.pdf}
\includegraphics[width=0.4\textwidth]{PL_spectrum_50_pct_8_16_-6_0.pdf}
\caption{{Top: ED spectrum at $\bar\nu = 2/5$ for $\kappa = 0.5$. The spin roton mode has vanished and the ground state is a nearly bright state.
Bottom: PL spectrum for the same system. Unlike the $SU(2)$ case, there is a peak visible even at $T = 0$, as in the $\bar\nu = 1/3$ case.} 
\label{fig:50_pct_2_5}
}
\end{figure}

As $\kappa$ is reduced strongly some of the conclusions of the $SU(2)$ symmetric case may stop being valid. One example is the vanishing of the spin roton gap at $\bar\nu = 2/5$ for $\kappa = 0.5$ as shown in Fig.~\ref{fig:50_pct_2_5}, in which case there is a peak visible even at $T=0$ like the $\bar\nu = 1/3$ case. Additionally, the main peak (the one adiabatically connected to the $SU(2)$ peak) becomes less prominent at higher temperatures. Further reduction in $\kappa$ may lead to a phase transition from the 332 type states considered in this paper to 331 type Halperin states with a single inter-band vortex \cite{Halperin83} or even a 330 type state with no inter-band vortex. We do not investigate these cases here, leaving them for a future study.  

\bibliographystyle{apsrev}
%
%

\end{document}